\documentclass[reqno,12pt]{amsart}
\usepackage{graphicx}
\usepackage{latexsym, amsfonts,mathrsfs, amsmath, amssymb, amscd, epsfig, ifthen, appendix}

\newtheorem{thm}{Theorem}[section]
\newtheorem{defn}[thm]{Definition}
\newtheorem{prop}[thm]{Proposition}
\newtheorem{lem}[thm]{Lemma}
\newtheorem{cor}[thm]{Corollary}
\newtheorem{rmk}{Remark}[section]

\numberwithin{equation}{section}

\newenvironment{pf}{{\noindent \it \bf Proof:}}{{\hfill$\Box$}\\}

\title[ ]{Asymptotic Behavior of Massless Dirac Waves in Schwarzschild geometry}

\date{}
\author[ ]{Joel Smoller$^*$} \thanks{$^*$Joel Smoller was supported in part by the NSF,  grant No. DMS-0603754.}
\author[ ]{Chunjing  Xie}
\address{Department of Mathematics, University of Michigan, 530 Church St. Ann Arbor, MI 48109 USA}
\email{smoller@umich.edu}
\email{cjxie@umich.edu}

\begin{document}

\def\ep{\varepsilon}
\def\s{\sigma}
\def\hK{\hat{K}}
\def\tK{\tilde{K}}
\def\ttK{\tilde{\tilde{K}}}

\def\ba{\begin{array}}
\def\ea{\end{array}}
\def\bma{\left(\begin{matrix}}
\def\ema{\end{matrix}\right)}
\def\be{\begin{equation}}
\def\ee{\end{equation}}

\def\nlz{{-, \lambda, 0}}
\def\nlo{{-, \lambda, 1}}
\def\nli{{-,\lambda, i}}
\def\nl{{-, \lambda}}
\def\pl{{+, \lambda}}
\def\pml{{\pm, \lambda}}
\def\lz{{\lambda, 0}}
\def\lo{{\lambda, 1}}
\def\li{{\lambda, i}}
\def\plz{{+, \lambda, 0}}
\def\plo{{+, \lambda, 1}}
\def\pli{{+,\lambda, i}}

\def\flp{f_\lambda^+}
\def\flm{f_\lambda^-}
\def\flpm{f_\lambda^\pm}
\def\Glp{\Gamma_\lambda^+}
\def\Glm{\Gamma_\lambda^-}
\def\Glpm{\Gamma_\lambda^\pm}
\def\qlp{q_\lambda^+}
\def\qlm{q_\lambda^-}
\def\qlpm{q_\lambda^\pm}

\def\bphi{\bar{\phi}}
\def\hphi{\hat{\phi}}
\def\tphi{\tilde{\phi}}
\def\vphi{\varphi}
\def\bvphi{\bar{\varphi}}
\def\hvphi{\hat{\varphi}}

\def\lur{\langle u\rangle}
\def\lxr{\langle x\rangle}
\def\lxpr{\langle x' \rangle}
\def\ltr{\langle t \rangle}
\def\ldr{\langle \cdot\rangle}

\def\mbN{\mathbb{N}}
\def\mbR{\mathbb{R}}
\def\mbC{\mathbb{C}}
\def\msS{\mathscr{S}}

\def\msHpml{\mathscr{H}_{\pm,\lambda}}
\def\msHpl{\mathscr{H}_{+,\lambda}}
\def\msHml{\mathscr{H}_{-,\lambda}}

\def\mcH{\mathcal{H}}
\def\mcK{\mathcal{K}}
\def\mfh{\mathfrak{h}}
\def\mstheta{\mathscr{\theta}}

\def\OC{O_{\mathbb{C}}}
\def\OI{O_{\mathfrak{I}}}
\def\mfr{\mathfrak{r}}
\def\mfs{\mathfrak{s}}

\def\Im{{\rm Im}}
\def\Re{{\rm Re}}

\def\mE{\mathcal{E}}
\def\mEm{\mathcal{E}_-}
\def\mMp{\mathcal{M}_+}
\def\mMm{\mathcal{M}_-}
\def\mEpm{\mathcal{E}_{\pm}}
\def\mMpm{\mathcal{M}_{\pm}}

\def\bma{\left(\begin{matrix}}
\def\ema{\end{matrix}\right)}

\maketitle

\begin{abstract}
In this paper, we show that massless Dirac
waves in the  Schwarzschild  geometry decay to zero at a rate $t^{-2\lambda}$, where $\lambda=1, 2, \cdots$ is the angular momentum.  Our technique  is to use Chandrasekhar's
separation of variables whereby  the Dirac equations split into two
sets of wave equations. For the first set, we show that the wave
decays as $t^{-2\lambda}$. For the second set, in general, the solutions  tend to some explicit profile at the rate $t^{-2\lambda}$.  The decay rate of solutions of Dirac equations is achieved by showing that the coefficient of the explicit profile  is exactly zero. The key ingredients in the proof of the decay rate of solutions for the first set of wave
equations  are an energy estimate used to show the absence of bound
states and zero energy resonance and the analysis of the spectral representation of the solutions.   The proof of asymptotic behavior
for the solutions of the second set of wave equations relies on
careful analysis of the Green's functions for time independent
Schr\"odinger equations associated with these wave equations.
 \end{abstract}

\section{Introduction and Main Results}
The vacuum Einstein equations
\begin{equation}
R_{ij}=0
\end{equation}
describe the evolution of spacetime with no sources, where $R_{ij}$ is the Ricci tensor of the Lorentzian metric $g_{ij}$.
The Schwarzschild solution of  the Einstein equations
describes the  geometry in the exterior  of a static and spherically symmetric black hole (cf. \cite{HawkingEllis}). It can be written as
\begin{equation*}
ds^2=\left(1-\frac{2M}{r}\right)dt^2-\frac{dr^2}{1-\frac{2M}{r}}-
r^2(d\theta^2+\sin^2\theta d\varphi^2),
\end{equation*}
where the coordinates $(t,r,\theta,\varphi)$ are in the range
\begin{equation*}
-\infty<t<\infty,\,\, 2M<r<\infty,\,\,
0\leq \theta\leq \pi,\,\, 0\leq \varphi<2\pi.
\end{equation*}

The stability of black holes is an important problem both
mathematically and physically. A first step is to
understand  the linearized problem. The study of linear stability of scalar waves in the Schwarzschild geometry
was initiated by Regge and Wheeler in \cite{ReggeW}, where the mode
stability for metric perturbations was discussed.  The
decay of the solution was proved in \cite{Twainy}.
Another approach uses the spherical symmetry in order to
produce an expansion into spherical modes, and to study the
corresponding ordinary differential equations. This was pursued by
Kronthaler in \cite{KronJMP, KronRate}, who  was
able to establish a decay rate in the spherical symmetric
case. Another analysis was carried out later in \cite{SofferPrice, DSwave}
for all spherical modes; they proved a $t^{-(2l+2)}$ local decay where $l=1, 2, \cdots$ is the angular momentum and improved the decay rate to $t^{-3}$ in the case $l=0$. The decay rate for solutions with general initial data was achieved in \cite{DafermosRCPAM, Tataru,SofferPoint}; see also \cite{DafermosRSimple,  BlueSI, BlueSspin2, BlueSter,  AlinhacSch, Luk, DafermosRPrice} for related results.

It is also of interest to study the linearized problem incorporating
various physical fields in the external region of a black hole, for
example, the radiation of gravitational, electromagnetic, and Dirac
waves.
The $L^{\infty}_{loc}$ decay for  massive  Dirac particles in Kerr-Newman geometry was proved  in \cite{FKSYDirac} .  Based on the spectral representation of the solutions for the Dirac equations in Kerr geometry,  the exact decay rate for  massive particles  with bounded angular momentum was obtained
in \cite{FKSYmassiveDirac}.
For electromagnetic  and gravitational radiation (corresponding to spin $s=1$ and $s=2$), numerical studies were given in \cite{TeuP}.   In \cite{WhitingMode},   a rigorous proof of mode stability  was obtained for the Teukolsky equation in the Kerr geometry. Recently,  in \cite{FinsterS},  Finster and Smoller obtained decay in $L_{loc}^{\infty}$ for the Teukolsky equation for spin  $s=1/2$, $1$ and $2$ (corresponding to Dirac equations, Maxwell equations, gravitational perturbations) in the Schwarzschild geometry. The decay of Maxwell's equations in the Schwarzschild geometry was studied in \cite{BlueMaxwell}.  In this paper, we consider  {\it massless} Dirac waves in the Schwarzschild geometry, and we prove that the solutions  decay to zero at a rate $t^{-2\lambda}$ where $\lambda=1, 2, \cdots$ is the angular momentum.

Introducing the Regge-Wheeler coordinate
\begin{equation}\label{RWcoordinate}
x=r+2M\ln(r-2M)-3M -2M\ln M,
\end{equation}
the massless Dirac equations in the Schwarzschild geometry (cf. \cite{Chandra}) can be written in Hamiltonian form as
\begin{equation}\label{DiracHeq}
i\frac{\partial}{\partial t}\Psi=H\Psi,
\end{equation}
where
\begin{equation*}
\begin{aligned}
H=\bma -\mE & 0 & 0 & 0\\
0 & \mE & 0 & 0\\
0 & 0 & \mE & 0\\
0 & 0 & 0 & -\mE \ema
+\bma 0 & -\mMp & 0 & 0\\
-\mMm & 0 & 0 & 0\\
0 & 0 & 0 & \mMp\\
0 & 0 & \mMm & 0 \ema,
\end{aligned}
\end{equation*}
and
\begin{equation*}
\begin{aligned}
\mE= i\frac{\partial}{\partial x},\,\,\,\,
\mMpm=
\frac{\sqrt{\Delta}}{r^2}\left(i\frac{\partial}{\partial
\theta}+i\frac{\cot
\theta}{2}\pm\frac{1}{\sin\theta}\frac{\partial}{\partial\varphi}\right),\,\, \,\, \Delta = r^2-2Mr.
\end{aligned}
\end{equation*}
The remarkable feature of these equations is their separability.
If we set
\begin{equation}\label{Psisep}
\begin{aligned}
\Psi_1=R_-(t,x)Y_-(\theta, \varphi),\,\, \Psi_2=R_+(t,x)Y_+(\theta, \varphi),\\
\Psi_3=R_+(t,x)Y_-(\theta, \varphi), \,\, \Psi_4=R_-(t,x)Y_+(\theta, \varphi),
\end{aligned}
\end{equation}
then $Y_-(\theta, \varphi)$ and $Y_+(\theta,\varphi)$ satisfy the pair of equations
\begin{equation}\label{SSharmonic}
\left(\frac{\partial}{\partial
\theta}+\frac{\cot
\theta}{2}-i\frac{1}{\sin\theta}\frac{\partial}{\partial\varphi}\right)Y_+=-\lambda Y_-, \,\,\,
\left(\frac{\partial}{\partial
\theta}+\frac{\cot
\theta}{2}+i\frac{1}{\sin\theta}\frac{\partial}{\partial\varphi}\right)Y_-=\lambda Y_+,
\end{equation}
and thus $Y_+$ and $Y_-$ are spin weighted spherical harmonics $-{}_{1/2}Y_{jm}$ and ${}_{-1/2}Y_{jm}$, respectively, where $j=\frac{1}{2}, \frac{3}{2}, \cdots$, see \cite{Goldberg, Spinweight}. Furthermore,  the eigenvalues $\lambda$ in (\ref{SSharmonic}) are of the form
\begin{equation*}
\lambda=j+\frac{1}{2}=1, 2, 3, \cdots
\end{equation*}
The radial functions  $R_-$ and $R_+$ solve the following first order equations depending on $\lambda$
\begin{equation}\label{Req}
(\frac{\partial}{\partial t}+\frac{\partial }{\partial x})R_\nl=\lambda \frac{\sqrt{\Delta}}{r^2}R_\pl,\,\, \quad \,\,
(\frac{\partial}{\partial t}-\frac{\partial }{\partial x})R_\pl=-\lambda \frac{\sqrt{\Delta}}{r^2}R_\nl.
\end{equation}
The main purpose of this paper is to study the Cauchy problem for (\ref{Req}) with initial data
\begin{equation}\label{RIC}
R_\nl(0, x)=g_\nl(x),\quad R_\pl(0, x)=g_\pl(x)
\end{equation}

Our main result, namely $t^{-2\lambda}$ decay rate of solutions for Dirac equations, is stated as follows.
\begin{thm}\label{thmR}
As $t\rightarrow
\infty$,  any solution $R_\pml$ of the problem (\ref{Req}) and
(\ref{RIC}), satisfies
\begin{equation}\label{Rest}
\begin{aligned}
& \|\ldr^{-3\lambda-1}R_\pml(t, \cdot) \|_{L^{\infty}(\mbR)}
 \leq C
t^{-2\lambda}\left\| \ldr^{3 \lambda+1} \sum_{i=0}^1\left(\left|\frac{d^i g_{\pl}}{dx^{i}}\right|+\left|\frac{d^i g_{\nl}}{dx^i}\right|\right)(\cdot)\right\|_{L^1(\mbR)},
\end{aligned}
\end{equation}
where we always denote  $\langle x \rangle = (1+|x|^2)^{1/2}$.
\end{thm}

We first give the main idea for the proof of Theorem \ref{thmR}.
Inspired by \cite{Chandra}, we introduce
\begin{equation*}
Z_\pl=R_\nl+R_\pl,\,\,\, Z_\nl=R_\nl-R_\pl.
\end{equation*}
Then $Z_\pml$ satisfies
\begin{equation}\label{Zeq}
\begin{aligned}
&\frac{\partial^2 Z_{\pml}}{\partial t^2}-\frac{\partial^2 Z_{\pml}}{\partial x^2}=-V_{\pml}Z_{\pml},
\end{aligned}
\end{equation}
where
\begin{equation}\label{defV}
V_{\pml}= \pm \frac{\partial}{\partial x}\left(\frac{\sqrt{\Delta}}{r^2}\lambda\right)+\frac{\Delta}{r^4}\lambda^2.
\end{equation}
Furthermore,  \eqref{Req} and \eqref{RIC} give the initial data for $Z_\pml$ as
\begin{equation}\label{DiracZIC}
Z_\pml(0, \cdot) =g_\nl\pm g_\pl\qquad \partial_t Z_\pml(0, \cdot) =\left(\lambda \frac{\sqrt{\Delta}}{r^2} \mp \frac{\partial}{\partial x} \right)(g_\nl \mp g_\pl)
\end{equation}

We have the following theorems on the solutions of the problem \eqref{Zeq} with  initial data
\begin{equation}\label{ZIC}
Z_\pml(0,x)=u_\pml(x), \,\quad \partial_tZ_\pml(0,x)=v_\pml(x).
\end{equation}
\begin{thm}\label{thmZp}
The solution $Z_\pl$ of the problem (\ref{Zeq}) and (\ref{ZIC}) satisfies for any integer $\alpha$ in the range $1 \leq \alpha\leq  2\lambda +1$,
\begin{equation*}
\|\langle \cdot \rangle^{-\alpha-\lambda }Z_{+, \lambda}(t, \cdot)\|_{L^{\infty}(\mbR)}\leq C t^{-\alpha+1}\|\langle \cdot \rangle^{\alpha+\lambda} (|u_\pl|+|u_\pl'|+|v_\pl|)(\cdot)\|_{L^1(\mbR)} .
\end{equation*}
If $\lambda=1$, then we have the improved estimate
\begin{equation*}
\|\langle \cdot \rangle^{-5 }Z_{+, 1}(t, \cdot)\|_{L^{\infty}(\mbR)}\leq C t^{-3}\|\langle \cdot \rangle^{5} (|u_{+,1}|+|u_{+,1}'|+|v_{+,1}|)(\cdot)\|_{L^1(\mbR)}.
\end{equation*}
\end{thm}
\bigskip

We also have the following asymptotic behavior for $Z_\nl$.
\begin{thm}\label{thmZm}
As $t\rightarrow
\infty$,  the solution $Z_\nl$ of the problem (\ref{Zeq}) and
(\ref{ZIC}) satisfies
\begin{equation}\label{estsol1}
\begin{aligned}
& \|\ldr^{- 3\lambda-1}(Z_\nl(t, \cdot) -D \hphi_\nlo(\cdot))\|_{L^{\infty}(\mbR)}\\
\leq & C
t^{-2\lambda}\left\|\ldr^{3 \lambda+1} \left(\sum_{i=0}^1 \left|\frac{d^i u_\nl}{dx^i} \right|+\left| v_\nl\right|\right) (\cdot)\right\|_{L^1(\mbR)},
\end{aligned}
\end{equation}
where
\begin{equation}\label{defB0thm}
\hphi_{\nlo}(x)=e^{-\lambda  \int_{-\infty}^x \frac{\sqrt{\Delta(x)}}{r^2(x)}dx},\quad\text{and}\quad
D = \int_\mbR  \hphi_\nlo(x) v_\nl(x) dx.
\end{equation}
\end{thm}

 Theorem \ref{thmR} is a direct consequence of Theorems \ref{thmZp} and \ref{thmZm}. Indeed,  when the initial data for $Z_\nl$ has the special form \eqref{DiracZIC}, the coefficient $D$ in Theorem \ref{thmZm} vanishes. This is illustrated by the following straightforward computation,
\begin{equation}\label{compD}
\begin{aligned}
D =&\int_\mbR \left(\frac{\sqrt{\Delta}}{r^2}(g_\pl +g_\nl) -\frac{\partial}{\partial x}(g_\nl+g_\pl) \right) \hphi_\nlo dx\\
=&\int_\mbR \left(\frac{\sqrt{\Delta}}{r^2} +\frac{\partial}{\partial x}\right)\hphi_\nlo(g_\pl+g_\nl)  dx\\
=  &0,
\end{aligned}
\end{equation}
where we use $\left(\frac{\sqrt{\Delta}}{r^2} +\frac{\partial}{\partial x}\right)\hphi_\nlo=0$. Thus we need only prove Theorems \ref{thmZp} and \ref{thmZm}.

Before proving these theorems, we  make some remarks and outline  the proof. We  represent the solutions of  (\ref{Zeq}) and (\ref{ZIC})  in terms of spectral measures of the associated Schr\"odinger operators
\begin{equation}\label{Schop}
\msHpml Z=-\frac{d^2}{dx^2} Z +V_\pml(x) Z.
\end{equation}
Since these spectral measures can be written explicitly in terms of Green's functions of the Schr\"odinger equation
\begin{equation}\label{Scheq}
\msHpml Z =\s^2 Z,
\end{equation}
where  $\s^2$ is called the {\it energy} of  solutions of the equation (\ref{Scheq}), a  key part of the paper is to study the Green's function of \eqref{Scheq}.

\begin{rmk}
The proof of Theorem \ref{thmZp} is inspired by \cite{SofferPrice, DSwave}. The key ingredient is to justify the absence of negative eigenvalues of $\msHpl$ and bounded zero energy solutions for $\msHpl$. This is achieved by  an energy estimate combined with the asymptotic behavior of zero energy solutions.
\end{rmk}

\begin{rmk}
The asymptotic profile for solutions of $Z_\pl$ is the bounded zero energy solution for $\msHpl$. This is the key difference from  this paper and \cite{SofferPrice, DSwave} in both analytic and phenomenological aspects. The perturbative solution for small energies constructed in \S \ref{secpersmall} gives the explicit representation of Green's function.
\end{rmk}

\begin{rmk}
For massless Dirac equations, the key observation is that the coefficient in front of the asymptotic profile (cf. \eqref{compD}) is exactly zero. Thus the solutions decay at a rate $t^{-2\lambda}$.
\end{rmk}

The organization of the paper is as follows. In \S \ref{secpre}, we give some preliminaries about  the potentials $V_\pm$ and the spectra of the Schr\"odinger operators \eqref{Schop} for which there are no negative eigenvalues. The zero energy solutions of the Schr\"odinger operators are investigated in \S \ref{seczerosol}, where we construct a  bounded zero energy solution for $\msHml$ and show the absence of bounded zero energy solution for $\msHpl$ using an energy estimate. In \S \ref{secrep}, we give a representation of the solution in terms of spectral measures and prove Theorem \ref{thmZp} using some techniques in \cite{SofferPrice, DSwave}. In \S \ref{secJost}, we review some basic properties of Jost solutions and Green's functions for equation \eqref{Scheq}, which are used in the representation of solutions.  \S \ref{secpersmall} is devoted to the construction of small energy solutions (i.e., solutions of equation \eqref{Scheq} with small $|\s|$)  in different spatial regions. The representation of Jost solutions by the solutions constructed in \S \ref{secpersmall} is given in \S \ref{secmatch}. In \S \ref{secasymp}, we first compute the Green's function explicitly and then prove the asymptotic behavior of  solutions of the wave equations with potential $V_\nl$.
In the appendix,  three lemmas on the existence and derivatives of solutions for Volterra integral equations are included for the convenience.

{\bf Notation:} For a given function $f$, we denote by $O(f(x))$ (respectively $\OC(f(x))$) a real (respectively complex) function that satisfies $|O(f(x))|\leq C|f(x)|$ (respectively $|\OC(f(x))|\leq C|f(x)|$)  for some constant $C$ in a specified range of $x$ which will be clear from the context. As in \cite{SofferPrice},  we say that $ O(x^{\gamma})$, $\gamma\in \mbR$, is {\it of symbol type}, if $\frac{d^k}{dx^k}O(x^\gamma)=O(x^{\gamma-k})$ for any $k\in \mbN_0=\{0\}\cup \mbN$. We denote generic constant by $C$, which may change throughout the paper.

\section{Preliminaries}\label{secpre}
We begin with the following lemma on relations between Regge-Wheeler coordinate and spherical coordinate, whose proof is straightfoward.
\begin{lem}\label{lemrx}
The function $x\mapsto r(x)$ which is the inverse function for
$x(r)$ defined  in \eqref{RWcoordinate},  has the asymptotic
behavior $r(x)=x(1+O(x^{-1+\ep}))$ as $x\rightarrow \infty$ for any $\ep\in (0,1)$ and
$r(x)=2M+M\sqrt{e}e^{x/(2M)} +O(e^{x/M})$ as $x\rightarrow-\infty$,
 where the $O-$terms are of symbol type.
\end{lem}

Using Lemma \ref{lemrx}, we have the following results on the asymptotic behavior of $V_{\pm}$ at $\pm\infty$.
\begin{prop}\label{corvasp}
The asymptotic behavior of $V_{\pml}$ is given by
\begin{equation}
\begin{aligned}
&V_{\pml}= O( \frac{\sqrt{2} \lambda}{16M^3 }e^{1/4} e^{\frac{x}{4M}}),\quad {\rm as}\,\, x\rightarrow -\infty,\\
&V_{\pml} = \frac{\lambda(\lambda\mp
1)}{x^2}(1+O(x^{-1+\ep})),\quad {\rm as}\,\, x\rightarrow  \infty,
\end{aligned}
\end{equation}
where the $O-$terms are of the symbol type. In addition, we have
\begin{equation}\label{asyVp1}
V_{+,1}=-\frac{M}{x^3}(1+O(x^{-1+\ep})),
\end{equation}
where the $O-$term is of the symbol type.
\end{prop}
\begin{pf}
By straightforward computations, one has
\begin{equation}\label{formvpml}
V_\pml=\pm \frac{(3M-r)\sqrt{\Delta}}{r^4}\lambda +\frac{\Delta}{r^4}\lambda^2.
\end{equation}
Using Lemma \ref{lemrx}, for $x\rightarrow -\infty$, we have
\begin{equation*}
\begin{aligned}
V_\pl= & \frac{\lambda(3M-r)\sqrt{\Delta}+\lambda^2 \Delta}{r^4}
\sim  \frac{\lambda M \sqrt{2M}\sqrt{M}e^{1/4} e^{\frac{x}{4M} }} {(2M)^4}
\sim   \frac{\sqrt{2} \lambda}{16M^3 }e^{1/4} e^{\frac{x}{4M}}.
\end{aligned}
\end{equation*}
For $x\rightarrow \infty$,  one has
\begin{equation*}
\begin{aligned}
V_\pl\sim  &\frac{-x^2 (1+O(x^{-1+\ep}))}{x^4}\lambda +\frac{\lambda^2 x^2(1+O(x^{-1+\ep}))}{x^4}\\
\sim  &\frac{(\lambda^2-\lambda)}{x^2}(1+O(x^{-1+\ep})).
\end{aligned}
\end{equation*}

Similarly, we can get the asymptotic behavior for $V_\nl$\,.

For \eqref{asyVp1}, we use \eqref{formvpml} to obtain as $r\to \infty$
\begin{equation}
\begin{aligned}
V_{+,1} = &\frac{3M}{r^4}-\frac{\sqrt{\Delta}}{r^3}+\frac{1}{r^2} -\frac{2M}{r^3}= \frac{3M}{r^4} -\frac{2M}{r^3} +\frac{1}{r^2} \left(1-\sqrt{1-\frac{2M}{r}}\right)\\
= & \frac{3M}{r^4} -\frac{2M}{r^3}+\frac{1}{r^2}\left(1-\left(1-\frac{M}{r}+O\left(\frac{1}{r^2}\right)\right)\right)\\
= & -\frac{M}{r^3}+O\left(\frac{1}{r^4}\right).
\end{aligned}
\end{equation}
Then the asymptotic behavior \eqref{asyVp1} follows from Lemma \ref{lemrx}\,.
\end{pf}

It is easy to see that $\msHpml$, defined by \eqref{Schop},
is a self adjoint operator on $L^2(\mbR)$ with domain $H^2(\mbR)$.
Proposition \ref{corvasp} shows that  $\msHpml$ is a compact perturbation of the self-adjoint operator $-\Delta$.
It follows from Weyl's theorem \cite{Teschl} that the essential  spectrum of $\msHpml$ is continuous and equals the absolute continuous spectrum,
\begin{equation}\label{essspectra}
\sigma_{ess}(\msHpml)= \sigma_{ac}(\msHpml)=[0,\infty).
\end{equation}

For the potential $V_\pl$, we have the following lemma.
\begin{lem}\label{lempvp}
If $\lambda\in \mbN$, then $V_\pl>0$ for $r\in (2M, \infty)$.
\end{lem}
\begin{pf}
It is easy to see from \eqref{formvpml} that $V_\pl>0$ if $r\in (2M, 3M)$.  Set $q(r)=\frac{3M-r}{\sqrt{\Delta}}$, then $V_\pl= \frac{[q(r)+\lambda]\lambda \Delta}{r^4}$.
A direct computation gives
\begin{equation*}
\frac{dq}{dr}=\frac{M(\frac{3M}{2}-r)}{\Delta^{3/2}}.
\end{equation*}
Hence $\frac{dq}{dr}<0$ when $r>\frac{3M}{2}$. Therefore, for $r\geq \frac{3M}{2}$, we have
\begin{equation*}
q(r)>\lim_{r\rightarrow \infty}q(r)=-1.
\end{equation*}
This implies that $V_\pl>0$ if $\lambda\geq 1$.
\end{pf}

The following lemma proves the absence of eigenvalues for $\msHpml$.

\begin{lem}\label{lemeig}
$\msHpml$ has no eigenvalues.
\end{lem}
\begin{pf}
That $\msHpl$ has no eigenvalues is a consequence of \eqref{essspectra} and Lemma \ref{lempvp}.

Let $Z_-$ be an eigenfunction  of $\msHml$ associated with an eigenvalue $\sigma<0$. That is,
\begin{equation}\label{eigeneqm}
\msHml  Z_-=\sigma Z_-.
\end{equation}
Set $Z=\lambda^2 \frac{\sqrt{\Delta}}{r^2} Z_-+\lambda Z_-'$. First, we claim that $Z$ is not zero. If $Z= 0$, then $Z_-=Ce^{-\int^x \lambda \frac{\sqrt{\Delta}}{r^2} dx}$.
By the direct computations, $\msHml Z_-=0$. Thus
\begin{equation*}
\msHml Z_--\sigma Z_- =-\sigma Z_- \neq 0.
\end{equation*}
This contradicts (\ref{eigeneqm}). Thus $Z$ is not zero.
Furthermore,  a straightforward computation shows that
\begin{equation}
\msHpl Z=\sigma Z.
\end{equation}
Hence $Z$ is an eigenfunction of $\msHpl$ corresponding to a negative eigenvalue $\sigma$. This contradicts the assertion that $\msHpl$ has no negative eigenvalue. Thus $\msHml$   has no eigenvalues.
\end{pf}

\section{Zero Energy Solutions}\label{seczerosol}

\subsection{Zero Energy Solutions for $\msHml$}
In this subsection, we investigate solutions of the equation $\msHml Z=0$.

Let
\begin{equation}\label{defhphinlz}
\begin{aligned}
\hphi_{\nlz}(x)=\hphi_{\nlo}(x)\int_0^x \frac{1}{\hphi_{\nlo}^2(y)}dy,
\quad \text{and}\quad
\hphi_{\nlo}(x)=e^{-\lambda  \int_{-\infty}^x \frac{\sqrt{\Delta(x)}}{r^2(x)}dx}.
\end{aligned}
\end{equation}
First, we have the following lemma on the properties of $\hphi_\nli$ ($i=0, 1$).
\begin{lem}\label{lemhphi}
$\hphi_\nlz$  and $\hphi_\nlo$ have the following properties.
\begin{enumerate}
\item  $\msHml \hphi_{\nli}=0$ for $i=0$, $1$.
\item  $Q_0 =\hphi_\nlz'(0)$ is positive and we can choose $x_0\in (0, 1)$ such that \begin{equation}\label{hphinlzs}
\begin{aligned}
&\frac{Q_0}{2} \leq \hphi_{\nlz}'(x)\leq 2Q_0 ,\,\,\,\, \text{if}\,\, -x_0<x<x_0.
\end{aligned}
\end{equation}
Moreover,   $\hphi_\nlz(x)>0$ for $ x>0$ and $\hphi_\nlz<0$ for $x<0$.
\item  $\hphi_{\nli}$ ($i=0, 1$) have the asymptotic behavior
\begin{equation}\label{hphipbehavior}
\begin{aligned}
&\hphi_\nlz(x)=\frac{1}{(2\lambda +1)B_0}x^{\lambda+1}(1+O(x^{-1+\ep})),\\
&\hphi_\nlo(x)=B_0x^{-\lambda}(1+O(x^{-1+\ep})),
\end{aligned}
\end{equation}
for $x>x_0$ for some positive constant $B_0>0$ and any $\ep \in (0, 1)$ and
\begin{equation}\label{hphinbehavior}
\begin{aligned}
&\hphi_\nlz(x)=x(1+\frac{B_1}{x} +O(e^{\frac{x}{4M}}) ),\,\, \,\, \hphi_\nlo(x)=1+O(e^{x/(4M)}),
\end{aligned}
\end{equation}
for some constant $B_1<0$ for $x<-x_0$, where the $O-$terms are of the symbol type.
\item Furthermore, the Wronskian of $\hphi_\nlz$ and $\hphi_\nlo$ satisfies
\begin{equation}\label{hphiW}
W(\hphi_\nlz,\hphi_\nlo)= \hphi_\nlz(x)\hphi_\nlo'(x)-\hphi_\nlz'(x)\hphi_\nlo(x)=-1.
\end{equation}
\end{enumerate}
\end{lem}
\begin{pf}
It is easy to check (1), (2) and (4) by straightforward computations. We need only  prove (3).

It follows from Lemma \ref{lemrx} that $r\to 2M$ and $\Delta=O(e^{\frac{x}{2M}})$ as $x\to -\infty$. Thus for $x\to -\infty$,
\begin{equation*}
\hphi_\nlo(x)=  e^{-\lambda \int_{-\infty}^x \frac{\sqrt{\Delta}}{r^2}dx} =1+ O(e^{\frac{x}{4M}})\quad\text{and}\quad   \hphi_\nlo'(x)= e^{-\lambda \int_{-\infty}^x \frac{\sqrt{\Delta}}{r^2}dx}  \frac{- \lambda\sqrt{\Delta}}{r^2} =O(e^{\frac{x}{4M}}).
\end{equation*}

For $x\to \infty$, it follows from Lemma \ref{lemrx} that
\begin{equation}
\begin{aligned}
&\frac{\sqrt{\Delta}}{r^2} -\frac{1}{x} =  \frac{1}{r} \sqrt{1-\frac{2M}{r}}-\frac{1}{x} \\
= & \frac{1}{x} (1+O(x^{-1+\ep}))\left(1-\frac{M}{x(1+O(x^{-1+\ep}))}\right) -\frac{1}{x}= O(x^{-2+\ep}).
\end{aligned}
\end{equation}
Therefore,
\[
\left|\int_1^\infty ( \frac{\sqrt{\Delta}}{r^2} -\frac{1}{x}) dx \right| < \infty
\quad
\text{and}
\quad
\int_x^\infty ( \frac{\sqrt{\Delta}}{r^2} -\frac{1}{x}) dx = O(x^{-1+\ep})
\]
for $x\to \infty$. Thus we have
\begin{equation}
\begin{aligned}
\int_{-\infty}^x \frac{\sqrt{\Delta}}{r^2} dx = &\int_{-\infty}^1 \frac{\sqrt{\Delta}}{r^2} dx + \int_1^{x} (\frac{\sqrt{\Delta}}{r^2} -\frac{1}{x}) dx  +\int_1^x \frac{1}{x} dx \\
= & \int_{-\infty}^1 \frac{\sqrt{\Delta}}{r^2} dx + \int_1^{\infty}( \frac{\sqrt{\Delta}}{r^2} -\frac{1}{x}) dx  +\int_1^x \frac{1}{x} dx -\int_x^\infty (\frac{\sqrt{\Delta}}{r^2} -\frac{1}{x}) dx.
\end{aligned}
\end{equation}
Let $B_0 =e^{-\lambda \int_{-\infty}^1 \frac{\sqrt{\Delta}}{r^2} dx  -\lambda \int_1^{\infty} (\frac{\sqrt{\Delta}}{r^2} -\frac{1}{x}) dx}$. Then as $x\to \infty$,
\begin{equation}
\hphi_\nlo(x) =B_0 e^{-\lambda \ln x +O(x^{-1+\ep})}=B_0x^{-\lambda}(1+O(x^{-1+\ep})).
\end{equation}

From \eqref{defhphinlz}, we have
\begin{equation*}
\hphi_\nlz'(x)=\hphi_\nlo'(x)\int_0^x \frac{1}{\hphi_\nlo^2(y) } dy+  \frac{1}{\hphi_\nlo(x) }.
\end{equation*}
This shows that for $x\to -\infty$, $\hphi_\nlz'(x) =1+O(e^{\frac{x}{6M}})$. Define $\mfr(x)= \frac{1}{\hphi_\nlo^2(x)}-1$. Then $0< \mfr(x) \leq Ce^{\frac{x}{4M}}$ for $x\in \mbR$.
For the asymptotic behavior of $\hphi_\nlz$, we have the following estimate
\begin{equation*}
\begin{aligned}
\hphi_\nlz(x)=&\hphi_\nlo(x) \int_0^x \frac{1}{\hphi_\nlo^2(y)} dy= (1+O(e^{\frac{x}{4M}}))\int_0^x 1+\mfr(y) dy \\
= & (1+O(e^{\frac{x}{4M}})) \left( x +\int_0^{-\infty} \mfr(y) dy +\int_{-\infty}^x  \mfr(y) dy \right)\\
= & (1+O(e^{\frac{x}{4M}})) \left( x +B_1 + O(e^{\frac{x}{4M}}) \right) \\
= &x(1+\frac{B_1}{x}+ O(e^{\frac{x}{4M}})) ,
\end{aligned}
\end{equation*}
where we denote $B_1=- \int_{-\infty}^0  \mfr(y) dy<0$. This finishes the proof of the asymptotic behavior of $\hphi_\nlz$.
\end{pf}

\subsection{Zero Energy Solutions for $\msHpl$}
In this subsection, we study solutions of the equation $\msHpl Z=0$. The main goal is to prove the absence of bounded solution for $\msHpl Z=0$.

\begin{lem}\label{lemphiasp}
There exist smooth functions $\hphi_\pli$ satisfying $\msHpl \hphi_\pli=0$ for $i=0$, $1$, with the asymptotic behavior
\begin{equation}\label{pphipbehavior}
\begin{aligned}
&\hphi_\plz(x)=\frac{1}{2\lambda -1}x^{\lambda}(1+O(x^{-1+\ep})),\,\, \,\,
\hphi_\plo(x)=x^{1-\lambda}(1+O(x^{-1+\ep})),
\end{aligned}
\end{equation}
as $x\rightarrow \infty$, for any $\ep \in (0, 1)$, where the $O-$terms are of symbol type.
Furthermore,
$\hphi_\nlz$ and $\hphi_\nlo$ satisfy
\begin{equation*}
W(\hphi_\plz,\hphi_\plo)= \hphi_\plz(x)\hphi_\plo'(x)-\hphi_\plz'(x)\hphi_\plo(x)=-1.
\end{equation*}
In addition, if $\lambda=1$, then
\begin{equation}\label{phphio}
\hphi_\plo'(x)=-\frac{M}{2} x^{-2}(1+O(x^{-1+\ep})).
\end{equation}
\end{lem}
\begin{pf}
The proof relies on the study on the integral equation for $\hphi_\plo/x^{1-\lambda}$. The property of $V_{\pl}$ obtained in Proposition \ref{corvasp} makes the proof for \eqref{pphipbehavior} similar to  that for \cite[Lemma 4.1]{SofferPrice}.

Now let's prove \eqref{phphio}. Indeed, if $\hphi_\plo(x)$ is a solution of $\msHpl Z=0$, then $\hphi_\plo(x)$ satisfies the integral equation
\begin{equation}
\hphi_\plo(x)=1 + \int_x^\infty (y-x) 	V_{+, 1}(y) \hphi_\plo(y) dy
\end{equation}
Therefore,  \eqref{asyVp1} and \eqref{pphipbehavior} give
\begin{equation}
\begin{aligned}
\hphi_\plo'(x)=&\int_x^\infty V_{+,1}(y) \hphi_\plo(y) dy\\
= & \int_x^\infty -M y^{-3}(1+O(y^{-1+\ep})) dy = -\frac{Mx^{-2}}{2} (1+O(x^{-1+\ep})).
\end{aligned}
\end{equation}
This proves \eqref{phphio}.
\end{pf}

For solutions of $\msHpl Z=0$ as $x\rightarrow-\infty$, we have
\begin{lem}\label{lempsiasp}
There exist smooth functions $\hat{\psi}_\pli$ satisfying $\msHpl \hat{\psi}_\pli=0$ for $i=0$, $1$ with the asymptotic behavior
\begin{equation}\label{psinasp}
\begin{aligned}
&\hat{\psi}_\plz(x)=x(1+O(e^{\frac{x}{8M}})),\,\,\,\,
\hat{\psi}_\plo(x)=1+O(e^{\frac{x}{8M}}),
\end{aligned}
\end{equation}
as $x\rightarrow -\infty$ with $O-$terms of symbol type. Furthermore, the Wronskian  of $\hat{\psi}_\plz$ and $\hat{\psi}_\plo$  is $W(\hat{\psi}_\plz,\hat{\psi}_\plo)=-1$.
\end{lem}
\begin{pf}
For $x\leq -1$, if $\hat{p}_\plz$ and $\psi_\plo$ satisfy the Volterra integral equations
\begin{equation*}
\hat{p}_\plz(x)=-\int_{-\infty}^x\left(\frac{y^2}{x} -y\right) V_{+,\lambda}(y) (1+\hat{p}_\plz(y)) dy
\end{equation*}
and
\begin{equation*}
\hat{\psi}_\plo(x)=1-\int_{-\infty}^x (y-x)V_{+, \lambda}(y)\hat{\psi}_\plo(y)dy,
\end{equation*}
then $\hat{\psi}_\plz(x)=x(1+\hat{p}_\plz(x))$ and $\hat{\psi}_\plo$  are solutions of the equation $\msHpl Z=0$.
Note that  for $a<-1$
\begin{equation*}
\sup_{x\in (y, a)} \left| \left(\frac{y^2}{x}-y\right) V_\pl(y) \right|\leq C e^{\frac{y}{8M}},
\end{equation*}
then applying Lemma \ref{lemvol} (in Appendix A) yields
\begin{equation*}
|\hat{p}_\plz(x)|\leq Ce^{\frac{x}{8M}}.
\end{equation*}
 Using Lemma \ref{lemderivative}, we get the asymptotic behavior of $\hat \psi_\plz$ in (\ref{psinasp}). Similarly, we can show the asymptotic behavior of $\hat \psi_\plo$ in (\ref{psinasp}).

 As $x\rightarrow -\infty$, we have
\begin{equation*}
\hat\psi_\plz \hat\psi_\plo'-\hat\psi_\plz'\hat\psi_\plo\rightarrow -1.
\end{equation*}
Since the Wronskian is a constant, we have $W(\hat{\psi}_\plz, \hat{\psi}_\plo) =-1$.
\end{pf}

Lemma \ref{lempsiasp} yields the following proposition
\begin{prop}
The solutions $\hphi_\pli$ for $i=0$, $1$, can be uniquely extended to $\mbR$ and we have $\hphi_\pli(x)=O(x)$ as $x\rightarrow -\infty$, where the $O-$term is of symbol type.
\end{prop}
\begin{pf}
Since $\hat{\phi}_\pli$ ($i=0, 1$) are solutions of a linear ODE with regular coefficients, we can extend the solution to the entire real line. As $x\rightarrow -\infty$, using Lemma \ref{lempsiasp},  there exist constants $c_{ij}$ such that
\begin{equation*}
\hat{\phi}_\pli(x)=c_{i0} \hat{\psi}_\plz(x)+c_{i1}\hat{\psi}_\plo(x).
\end{equation*}
Then Lemma \ref{lempsiasp} implies the proposition.
\end{pf}

The important consequence of Lemmas \ref{lemphiasp} and \ref{lempsiasp} is the following:
\begin{lem}\label{lemnobdd}
Any bounded  solution of $\msHpl Z=0$ on $\mbR$ must be zero.
\end{lem}
\begin{pf} We prove the lemma by contradiction.

Suppose that $Z$ is a nonzero  bounded solution for $\msHpl Z=0$.
It follows from Lemma \ref{lempsiasp} that  $Z(x)\rightarrow C_1$ for some constant $C_1$ and $Z'(x)\rightarrow 0$ as $x\rightarrow -\infty$. Furthermore, if $\lambda \geq 2$, then Lemma \ref{lemphiasp} implies that $Z(x) \rightarrow 0$ and $Z'(x) \rightarrow 0$ as $x\rightarrow \infty$; if $\lambda =1$, then $Z(x)\to C_2$ for some constant $C_2$ and $Z'(x) \to 0$. Multiplying the equation $\msHpl Z=0$ by $Z$ and integrating over $\mbR$ yield
\begin{equation*}
\begin{aligned}
0= \int_{\mbR} (-Z''+V_\pl Z)Z dx = - ZZ'\big|_{-\infty}^{\infty} + \int_{\mbR}(|Z'|^2+V_\pl Z^2)dx.
\end{aligned}
\end{equation*}
Using the asymptotic behavior of $Z$ at $\pm\infty$ gives
\begin{equation*}
 \int_{\mbR}(|Z'|^2+V_\pl Z^2)dx=0.
 \end{equation*}
Since $V_\pl >0$ for $x\in \mbR$, we conclude that $Z=0$.
\end{pf}

\section{Representation of the Solutions and Proof of Theorem \ref{thmZp}}\label{secrep}
Since $\sigma(\msHpml)=\sigma_{ac}(\msHpml)=(0, \infty)$, the solution of the problem
\eqref{Zeq} and \eqref{ZIC} can be written as
\begin{equation}
Z_\pml(t)=\cos(t\sqrt{\msHpml})u_\pml+\frac{\sin(t\sqrt{\msHpml})}{\sqrt{\msHpml}}v_\pml,
\end{equation}
where
\begin{equation}\label{cossol}
\cos(t\sqrt{\msHpml})u_\pml=-\frac{2}{\pi} \int_0^\infty\int_\mbR \s \cos(t\s) \Im[G_\pml(x, x', \s)] u_\pml(x')dx'd\s
\end{equation}
and
\begin{equation}\label{sinsol}
\frac{\sin(t\sqrt{\msHpml})}{\sqrt{\msHpml}} v_\pml=-\frac{2}{\pi} \int_0^\infty\int_\mbR \sin(t\s) \Im[G_\pml(x, x', \s)] v_\pml(x')dx'd\s,
\end{equation}
and $G_\pml(x, x', \s)$ is the Green's function of the equation $\msHpml Z=\s^2
Z$, see \cite{SofferPrice}.

If $\lambda\geq 1$, then  using Lemma \ref{lemnobdd} and the techniques in \cite{SofferPrice}  for $1 \leq \alpha\leq 2\lambda+1$ gives
\begin{equation*}
\|\langle \cdot \rangle^{-\alpha-\lambda}Z_\pl(t, \cdot)\|_{L^{\infty}(\mbR)}\leq
C_{\lambda, \alpha}t^{-\alpha+1}\|\langle \cdot \rangle^{\alpha+\lambda}  (\sum_{i=0}^1|\frac{d^i u_\pl}{dx^i}| +|v_\pl|)(\cdot)\|_{L^1(\mbR)} .
\end{equation*}
In case of $\lambda=1$, since $V_{+, 1}$ satisfies \eqref{asyVp1}, then one can use Lemma \ref{lemnobdd} and \cite[Theorem 1.1]{DSwave} to obtain
\begin{equation*}
\|\langle \cdot \rangle^{-5}Z_\pl(t, \cdot) \|_{L^{\infty}(\mbR)}\leq C
t^{- 3} \|\langle \cdot \rangle^{5} (\sum_{i=0}^1|\frac{d^i u_\pl}{dx^i}| +|v_\pl|)(\cdot)\|_{L^1(\mbR)} .
\end{equation*}

This proves Theorem \ref{thmZp}. \qed

%\newpage

\section{Basic Properties of  Jost Solutions and Green's Function}\label{secJost}
In this section,  we discuss the Jost solutions and Green's function for the equation
\begin{equation}\label{eqmsHmls}
\msHml \phi =\s^2 \phi.
\end{equation}
\subsection{Existence of the Jost Solutions}
Recall that the Jost solutions $f_\lambda^\pm(\cdot, \s)$ are defined by $\msHml f_\lambda^\pm =\s^2 f_\lambda^\pm(\cdot, \s)$ with the asymptotic behavior $f_\lambda^\pm(x, \s) \sim e^{\pm i\s x}$ as $x\rightarrow \pm \infty$.
We begin with the following lemma on the existence of Jost solutions for any  $ \s \in \overline{\mbC}_+\backslash \{0\}$ where $\mbC_+=\{\s \in \mbC: {\rm Im} \s >0\}$, whose proof is similar to \cite[Lemma 3.1]{SofferPrice}.
\begin{lem}\label{LemJost}
For every $\s \in \bar\mbC_+\backslash \{0\}$ there exist smooth functions $f_{\lambda}^{\pm}(\cdot, \s)$ satisfying \eqref{eqmsHmls}
and $f_\lambda^{\pm}(x, \s)\sim e^{\pm i \s x}$ and $\partial_x f_\lambda^{\pm}(x, \s)\sim i\sigma e^{\pm i \s x}$for $x\rightarrow \pm \infty$. Furthermore, for every $x\in \mbR$, the functions $f_\lambda^{\pm}(x, \cdot)$ and $\partial_x f_\lambda^{\pm}(x, \cdot)$ are continuous functions in $\bar\mbC_+ \backslash \{0\}$.
\end{lem}

\subsection{The Jost Solutions at Large Energies} In order to estimate (\ref{cossol}) and (\ref{sinsol}) for the contributions from large energies, we need the behavior of the Jost solutions for large $\s$. Let
\begin{equation}\label{defGlpm}
\Glpm(x, \s) =e^{\mp i \s x} \flpm(x, \s)\quad \text{and}\quad \quad \qlpm(x,\s)=\Glpm(x,\s)-1.
\end{equation}
\begin{lem}\label{lembigs}
Let $\s_0>0$. Then, for $k$, $l\in \mbN$, the function $q_\lambda^+(\cdot, \s)$ satisfies the estimate
\begin{equation}
|\partial_x^k \partial_\s^l q_\lambda^+(x, \s)|\leq C \langle x\rangle^{-1-k} \s^{-1-l}
\end{equation}
for all $\s\geq \s_0$ and all $x\geq 0$. The same bound for $q_\lambda^-(x, \s)$ holds if $x\leq 0$.
\end{lem}
The proof is based on the estimate for the Volterra integral equation for $\qlpm$, which is quite similar to that in \cite[Lemma 9.1]{SofferPrice}.

\subsection{The Jost Solution $f_\lambda^-(\cdot, \s)$ for Small $\s$}
In this subsection, we investigate the Jost solution $\flm(x, \s)$ as $\s\rightarrow 0$.
\begin{lem}\label{lemflm}
Let $a\in \mbR$ and $\s_0>0$. Then the Jost solution $f_\lambda^-(x, \s)=e^{-i\s x} (1+ q_\lambda^-(x, \s))$ exists for all $\s\in [-\s_0, \s_0]$ and satisfies the bound
\begin{equation}\label{estqlm}
|\partial_{\s}^{m}\partial_x^k q_\lambda^-(x, \s)|\leq C(m, k, a) (e^{\frac{x}{2M}} + e^{\frac{x}{8M}})
\end{equation}
for all $x\in (-\infty, a]$ and all $\s \in [-\s_0, \s_0]$.
\end{lem}
\begin{pf}
If $q_\lambda^-(x, \lambda)$ satisfies the Volterra equation
\begin{equation}\label{eqqlm}
q_\lambda^-(x, \s) =\int_{-\infty}^x K(x, y, \s) (1+q_\lambda^-(y, \s))d y,
\end{equation}
where
\begin{equation*}
K(x, y, \s) =\frac{1}{2i\s}(e^{2i\s (x-y)}-1) V_{\nl}(y)=(x-y)V_\nl(y)\int_0^1 e^{2i \s (x-y) \theta}d\theta,
\end{equation*}
then $\flm(x, \s)=e^{-ix\s} (1+\qlm(x, \s))$ is a Jost solution of equation \eqref{eqmsHmls}. On $y\leq x\leq a$, we have
\begin{equation}\label{Kest}
|\partial_\s^l K(x, y,\s) |\leq C(l, a) |y-x|^{l+1}|V_{\nl}(y)|.
\end{equation}
Using Proposition \ref{corvasp}, Lemmas \ref{lemvol} and \ref{lemparameter} yield
\begin{equation}\label{qest}
\|\partial_\s^l q_{\lambda}^-(\cdot, \s)\|_{L^\infty(-\infty, a)}\leq C(l, a),
\end{equation}
for all $\s\in [-\s_0, \s_0]$ since $V_{\nl}(y)$ decays
exponentially as $y\rightarrow -\infty$. Note that for $y\in \mbR$,
$|V_\nl(y)|\leq C  e^{\frac{y}{4M}}$, so
\begin{equation*}
\begin{aligned}
|\qlm(x, \s)|=&\left|\int_{-\infty}^x K(x, y, \s) (1+q_\lambda^-(y, \s))d y\right| \\
\leq & C\int_{-\infty}^x  |V_\nl(y)|\cdot|y-x|\cdot  |(1+q_\lambda^-(y, \s))| dy\\
\leq & C \int_{-\infty}^x |V_\nl(y)|\cdot |y-x| dy\\
\leq & C (e^{\frac{x}{3M}} + e^{\frac{x}{6M}}),
\end{aligned}
\end{equation*}
where we used \eqref{Kest} and \eqref{qest} in the first and second inequality, respectively. Similarly, if we differentiate \eqref{eqqlm} with respect to $x$ and $\s$, we can show \eqref{estqlm} using Lemmas \ref{lemderivative} and \ref{lemparameter}.
\end{pf}

%\newpage
\subsection{Green's Function and Wronskians}
Standard techniques give that Green's function for \eqref{eqmsHmls} has the form
\begin{equation}\label{defGreen}
G(x,x', \s) =\frac{f_\lambda^-(x', \s) f_\lambda^+(x, \s)
\Theta(x-x')+f_\lambda^-(x, \s) f_\lambda^+(x', \s) \Theta(x'-x)}
{W(f_\lambda^-(\cdot, \s), f_\lambda^+(\cdot, \s))}
\end{equation}
for $\Im \s>0$, where $\Theta$ denotes the Heaviside function
\begin{equation*}
\Theta(x)=\left\{
\begin{aligned}
&1\quad \text{if}\,\, x>0,\\
&0\quad \text{if}\,\, x<0,
\end{aligned}
\right.
\end{equation*}
and $W(f_\lambda^-(\cdot, \s), f_\lambda^+(\cdot, \s))$ is the
Wronskian of $f_\lambda^-$ and $f_\lambda^+$. For ease in notation, we denote $W(\s)=W(\flm(\cdot, \s), \flp(\cdot, \s))$.
\begin{lem}\label{lemwron}
The Wronskian of the Jost solutions,  $W(\s)$, satisfies
\begin{equation}
|W(\s)| \geq 2\s
\end{equation}
for $\s>0$.
\end{lem}
\begin{pf}
The proof is quite similar to that for \cite[Lemma 3.2]{SofferPrice}.

It follows from Lemma \ref{LemJost} that
\begin{equation}
W(\overline{\flp}, \flp )= W(f_\lambda^-, \overline{\flm}) =2i \s
\end{equation}
which shows that $f_\lambda^+(\cdot, \s)$ and $\overline{f_\lambda^+(\cdot, \s)}$ are linearly independent for $\s>0$. Hence, for any Jost solution $\flm$, there exist $a(\s)$ and $b(\s)$ such that
\begin{equation}\label{trancoe1}
\flm (x, \s) =a(\s) \flp (x, \s) +b(\s) \overline{f_\lambda^+(\cdot, \s)}.
\end{equation}
We conclude that
\begin{equation*}
\begin{aligned}
2i \s = & W(f_\lambda^-, \overline{\flm})(\s)=W(af_\lambda^++b\overline{f_\lambda^+}, \overline{af_\lambda^+}+\bar{b}f_\lambda^+)(\s)\\
= & -2i \s |a(\s)|^2 +2i\s |b(\s)|^2,
\end{aligned}
\end{equation*}
which implies
\begin{equation}\label{tranrelation}
|b(\s)|^2-|a(\s)|^2=1.
\end{equation}
Thus $|b(\s)|\geq 1$. However,
\begin{equation}\label{tranbW}
W(\s) =W(f_\lambda^-, f_\lambda^+)(\s) =W(af_\lambda^+ + b\overline{f_\lambda^+}, f_\lambda^+)=2i \s b(\s),
\end{equation}
and thus $|W(\s)|\geq 2\s$.
\end{pf}

%\newpage

\section{Perturbative Solutions for $|\s|$ Small}\label{secpersmall}

\subsection{Construction of Perturbative Solutions for $|x\s|$ Small}
With the aid of Lemma \ref{lemhphi}, we can construct solutions of  equation  (\ref{eqmsHmls}).

\begin{lem}\label{lemh}
There exist solutions of equation (\ref{eqmsHmls}),  $\phi_{\nli}(x, \s)$ ($i=0, 1$), which have the form
\begin{equation*}
\phi_{\nli}(x, \s)=\hphi_{\nli}(x)(1+h_{\nli}(x, \s))
\end{equation*}
and satisfy $W(\phi_\nlz(\cdot, \s), \phi_\nlo(\cdot, \s))=-1$.
Furthermore, there are constants $\s_1$, $\delta\in (0, 1)$ such that
\begin{equation*}
|h_\nlz(x, \s)|\leq Cx^2\s^2,\quad \text{for}\,\, x\in [-\delta \s^{-1}, \delta \s^{-1}],
\quad \text{and}
\quad |h_\nlo(x, \s)|\leq C|x|  \lxr \s^2,
\end{equation*}
for  $x\in [-\delta \s^{-1}, \delta \s^{-1}]$,  $\s\in (0,\s_1)$ and $\delta \in (0, \delta_1)$, where $C$ is a constant independent of $\s$ and $\delta$.
\end{lem}
\begin{pf}
It follows from Lemma  \ref{lemhphi} that $\hphi_{\nlz}(x)\neq 0$ for $x\neq 0$.
Let $h_{\nlz}(x, \s)$ satisfy the Volterra integral equation
\begin{equation}\label{eqhnlz}
h_\nlz(x, \s)= \int_{0}^x \hK(x, y, \s) (1+h_\nlz(y,\s)) dy,
\end{equation}
with the kernel function
\begin{equation}\label{defhK}
\hK(x,y, \s) = -\s^2  \left(\hphi_\nlz(y)\hphi_\nlo(y) -\frac{\hphi_\nlo(x)}{\hphi_\nlz(x)}\hphi_\nlz^2(y)\right).
\end{equation}
Then $\phi_\nlz(x,\sigma)=\hphi_\nlz(x)(1+h_\nlz(x, \s))$ is a solution of (\ref{eqmsHmls}). If $y>0$, Lemma \ref{lemhphi} implies that
\begin{equation}\label{K2est}
\sup_{x\in(y, \delta \s^{-1})}|\hK(x, y, \s)|\leq C\s^2 y.
\end{equation}
Thus
\begin{equation*}
\int_0^{\delta\s^{-1}} \sup_{x\in(y, \delta \s^{-1})}|\hK(x, y, \s)|dy\leq C\delta^2.
\end{equation*}
It follows from Lemma \ref{lemvol}  that $|h_{\nlz}(x, \s)|\leq C$ for $x\in [0, \delta \s^{-1}]$.
Combining \eqref{K2est} and the integral equation \eqref{eqhnlz} gives
\begin{equation}\label{esthnlop}
|h_\nlz(x, \s)|\leq C \int_0^x C \s^2 y dy \leq C \s^2x^2 \qquad \text{for}\,\, x\in [0, \delta \s^{-1}].
\end{equation}
If $y<0$, then
\begin{equation*}
\sup_{x\in(-\delta \s^{-1}, y)}|\hK(x, y, \s)|\leq C\s^2 |y|.
\end{equation*}
It follows from  Lemma \ref{lemvol}  and  (\ref{eqhnlz}) that
\begin{equation}\label{esthnlon}
|h_{\nlo}(x, \s)|\leq Cx^2\s^2\qquad \text{for}\,\, x\in [-\delta \s^{-1}, 0].
\end{equation}
Choose $\delta_1\in (0, 1)$ such that $|h_\nlz(x, \s)|\leq C \delta_1^2\leq 1/2$ for $x\in [-\delta \s^{-1}, \delta \s^{-1}]$ and $\delta \in (0, \delta_1)$.

Define
\begin{equation*}
\begin{aligned}
\phi_\nlo(x, \s)=& \phi_\nlz(x, \s)\left(\int_{\delta\s^{-1}}^{\infty} \frac{1}{\hphi_{\nlz}^2(y)} dy+ \int_x^{\delta\s^{-1}} \frac{1}{\phi_\nlz^2}(y, \s) dy \right).
\end{aligned}
\end{equation*}
 A direct computation using \eqref{hphinlzs} and \eqref{hphipbehavior} shows   that $\phi_\nlo(x, \s)$ is well-defined for $x\in (-\delta \s^{-1}, \delta \s^{-1})$.
Then  $W(\phi_\nlz,\phi_\nlo)=-1$. Hence $\phi_\nlo$ is also a solution of (\ref{eqmsHmls}).
Let
\begin{equation}\label{defth0}
h_{\nlo}(x, \s) = \frac{\phi_\nlo(x, \s)}{\hphi_{\nlo}(x)}-1,
\quad \text{and}\quad
\tilde{h}_{\nlz}(x,\s) = (1+h_{\nlz}(x, \s))^{-2} -1.
\end{equation}
Using (\ref{esthnlop}) and (\ref{esthnlon}), we obtain
\begin{equation*}
|\tilde{h}_{\nlz}(x, \s)|\leq C\s^2 x^2\qquad \text{for}\,\, x\in [-\delta \s^{-1}, \delta \s^{-1}].
\end{equation*}

Note that \eqref{hphiW} also implies that $\left(\frac{\hphi_\nlo}{\hphi_\nlz}\right)'=-\frac{1}{\hphi_\nlz^2}$, so
\begin{equation*}
\hphi_{\nlo}(x)=\hphi_{\nlz}(x)\int_{x}^{\infty}\hphi_{\nlz}^{-2}(y) dy.
\end{equation*}
Therefore,
\begin{equation}\label{defhnlo}
\begin{aligned}
&h_{\nlo}(x, \s) = \frac{\hphi_{\nlz}(x) (1+h_{\nlz}(x, \s))}{\hphi_{\nlo}(x)}\left[ \int_{\delta \s^{-1}}^{\infty} \hphi_{\nlz}^{-2}(y) dy  \right.\\
& \quad\quad + \int_x^{\delta \s^{-1}} \hphi_{\nlz}^{-2}(y) dy +\left. \int_x^{\delta \s^{-1}} \hphi_{\nlz}^{-2}(y) \tilde{h}_{\nlz}(y, \s) dy \right] -1\\
& \quad =  h_{\nlz}(x, \s)+ \frac{\hphi_{\nlz}(x) (1+h_{\nlz}(x, \s))}{\hphi_{\nlo}(x)}  \int_x^{\delta \s^{-1}} \hphi_{\nlz}^{-2}(y) \tilde{h}_{\nlz}(y, \s) dy.
\end{aligned}
\end{equation}
For $x\in [x_0, \delta \s^{-1}]$, we have
\begin{equation*}
\begin{aligned}
|h_\nlo(x, \s) |\leq & C\s^2 x^2 + Cx^{2\lambda+1} \int_x^{\delta \s^{-1}} Cy^{-2\lambda-2} \s^2 y^2 dy \leq  C\s^2 x^2.
\end{aligned}
\end{equation*}
For $x\in [0, x_0]$, it follows from (\ref{hphinlzs}) that
\begin{equation}\label{esthatz}
\begin{aligned}
& \left|\int_x^{\delta \s^{-1}} \hphi_{\nlz}^{-2}(y) \tilde{h}_\nlz(y, \s) dy\right| \\
\leq & \left|\int_{x}^{x_0} \hphi_{\nlz}^{-2}(y) \tilde{h}_\nlz(y, \s) dy \right|+  \left|\int_{x_0}^{\delta \s^{-1}} \hphi_{\nlz}^{-2}(y) \tilde{h}_\nlz(y, \s) dy\right|\\
\leq & \int_0^{x_0} C y^{-2} \s^2 y^2 dy +\int_{x_0}^{\delta \s^{-1}} C y^{-2\lambda -2} \s^2 y^2 dy
\leq  C \s^2.
\end{aligned}
\end{equation}
Define $a_0 = \int_0^{\delta \s^{-1}} \hphi_{\nlz}^{-2}(y) \tilde{h}_\nlz(y, \s) dy$. Then (\ref{esthatz}) shows
\begin{equation}\label{esta0}
|a_0|\leq C\s^2.
\end{equation}
For $x\leq -x_0$, using (\ref{hphinlzs}) and (\ref{esta0}), we have
\begin{equation*}
\begin{aligned}
|h_\nlo(x,\s)|= & |h_\nlz(x,\s)| +\left|Cx\left(a_0+\int_{-x_0}^0  \hphi_\nlz^{-2}(y) \tilde{h}(y, \s)dy \right.\right.\\
&\left.\left.+ \int_x^{-x_0}  \hphi_\nlz^{-2}(y) \tilde{h}(y, \s)dy\right)\right|\\
\leq & C \s^2 |x|^2 +C|x|\s^2 \leq  C\s^2 |x|\lxr.
\end{aligned}
\end{equation*}
The above estimate also shows that for $x\in [-x_0, x_0]$
\begin{equation*}
|h_\nlo(x, \s)|\leq C|x|\lxr \s^2.
\end{equation*}

This finishes the proof of the lemma.
\end{pf}

%\newpage
\subsection{Derivative Estimates for the Perturbative Solutions for Small $\mathbf{|x\s|}$}
It is easy to see that $\phi_\nlz(x, \s) =\hphi_\nlz(x)(1+h_\nlz(x, \s))$ with $h_\nlz(x, \s)$ constructed in Lemma \ref{lemh}
satisfies equation (\ref{eqmsHmls}) and initial data
\begin{equation}
\phi_\nlz(0, \s) =\hphi_\nlz(0) =0, \quad \partial_x \phi_\nlz(0, \s) =\hphi_\nlz'(0)= Q_0.
\end{equation}
Therefore, $\phi_\nlz(x, \s)$ satisfies integral equation
\begin{equation}\label{eqphi610}
\phi_\nlz(x, \s) =Q_0 x +\int_0^x ( V_\nl(y) -\s^2)\cdot(x-y) \phi_\nlz(y, \s) dy.
\end{equation}
We first have the following estimate for $\phi_\nlz(x, \s)$.
\begin{lem}\label{lem62}
For $x \in [-x_0, x_0]$ where $x_0$ is the constant appeared in Lemma \ref{lemhphi}, we have
\begin{equation}\label{estphig6}
|\partial_\s^i \partial_x^j \phi_\nlz(x, \s)|\leq C
\end{equation}
for $i\in \mbN_0$ and $j\in \mbN_0$ and
\begin{equation}\label{estphi611}
|\partial_\s^{2k} \phi_\nlz(x, \s)|\leq C |x|,\quad \text{and}\quad   |\partial_\s^{2k+1} \phi_\nlz(x, \s)|\leq C |x| \s
\end{equation}
for $k \in \mbN_0$.
\end{lem}
\begin{pf}
\eqref{estphig6} is the direct consequence of Lemmas \ref{lemvol}-\ref{lemparameter}.
We prove \eqref{estphi611} by  induction.

Obviously,  equation \eqref{eqphi610} is a Volterra integral equation. Using Lemma \ref{lemvol}, we have $|\hphi_\nlz(x, \s) |\leq C|x|$ for $|x|\leq |x_0|\leq 1$.  Differentiating \eqref{eqphi610} with respect to $\s$ yields
\begin{equation*}
\begin{aligned}
\frac{\partial\phi_\nlz}{\partial \s} (x, \s)=& \int_0^x -2\s (x-y) \phi_\nlz(y, \s) dy\\
 & + \int_0^x ( V_\nl(y) -\s^2)\cdot(x-y) \frac{\partial \phi_\nlz}{\partial \s}(y, \s) dy.
\end{aligned}
\end{equation*}
It follows from Lemma \ref{lemvol} that $|\partial_\s\phi_\nlz(x, \s)|\leq C\s |x|$. Therefore, \eqref{estphi611} holds for $k=0$.

Now suppose that the estimate \eqref{estphi611}  holds for $k =0, \cdots, l-1$.  Differentiating \eqref{eqphi610} with respect to $\s$ $2l$-times gives
\begin{equation}
\begin{aligned}
\partial_\s^{2l}\phi_\nlz(x, \s) = &\int_0^x -2l(2l-1) (x-y) \partial_\s^{2l-2}\phi_\nlz(y, \s) dy\\
& + \int_0^x -4l \s (x-y) \partial_\s^{2l-1}\phi_\nlz(y, \s) dy\\
&+\int_0^x (V_\nl(y)-\s^2) (x-y) \partial_\s^{2l}\phi_\nlz(y, \s) dy.
\end{aligned}
\end{equation}
Since \eqref{estphi611} holds for $k=l-1$, we have
\[
|\int_0^x -2l(2l-1) (x-y) \partial_\s^{2l-2}\phi_\nlz(y, \s) -4l \s (x-y) \partial_\s^{2l-1}\phi_\nlz(y, \s) dy|\le C|x|.
 \]
Thus $|\partial_\s^{2l} \phi_\nlz(x, \s)|\leq C|x|$ follows from Lemma \ref{lemvol}.

Similarly, differentiating
\eqref{eqphi610} with respect to $\s$ \,\,  $(2l+1)$-times gives
\begin{equation}
\begin{aligned}
\partial_\s^{2l+1}\phi_\nlz(x, \s) = &\int_0^x -2l(2l+1) (x-y) \partial_\s^{2l-1}\phi_\nlz(y, \s) dy\\
& + \int_0^x -2(2l+1) \s (x-y) \partial_\s^{2l}\phi_\nlz(y, \s) dy\\
&+\int_0^x (V_\nl(y)-\s^2) (x-y) \partial_\s^{2l+1}\phi_\nlz(y, \s) dy.
\end{aligned}
\end{equation}
Since $|\partial_\s^{2l}\phi_\nlz(y, \s)|\leq C|y|$ for $y \in [-x_0, x_0]$,
\[
|\int_0^x -2l(2l+1) (x-y) \partial_\s^{2l-1}\phi_\nlz(y, \s)  -2(2l+1) \s (x-y) \partial_\s^{2l}\phi_\nlz(y, \s) dy|\leq C\s |x|.
\]
Therefore,  $|\partial_\s^{2l+1} \phi_\nlz(x, \s)|\leq C|x|\s$. This implies \eqref{estphi611} holds for $k=l$.
\end{pf}

For $x\in [-x_0, x_0]$, we  rewrite $h_\nlz(x, \s)$ as
\begin{equation}\label{h614}
\begin{aligned}
h_\nlz(x, \s)= &\frac{\phi_\nlz(x, \s)}{\hphi_\nlz(x)} -1=  \frac{\phi_\nlz(x, \s)/x}{\hphi_\nlz(x)/x} -1\\
= &\frac{\int_0^1 \int_0^{\theta x} ((V_\nl (y) -\s^2) \phi_\nlz(y, \s)-V_\nl(y)\hphi_\nlz(y))  dyd\theta}{\int_0^1 \hphi_\nlz'(\theta x) d\theta},
\end{aligned}
\end{equation}
where we used $\hphi_\nlz''(x) =V_\nl(x) \hphi_\nlz(x)$.
Using Lemma \ref{lem62}, we have the following estimates for $h_\nlz(x, \s)$.
\begin{lem}\label{lemh63}
Let $x_0$ be the constant from Lemma \ref{lemhphi}. For  $x\in [-x_0, x_0]$,  we have
\begin{equation}\label{esthgen}
|\partial_\s^k \partial_\s^l h_\nlz(x, s)|\leq C
\end{equation}
for $k$, $l\in \mbN_0$; and
\begin{equation}\label{esth615}
\sum_{i=0}^1| \partial_\s^{2k} \partial_x^i h_\nlz(x, \s)| \leq C |x|,\quad \text{and}\quad   \sum_{i=0}^1|\partial_\s^{2k+1}\partial_x^i h_\nlz(x, \s)| \leq C |x| \s
\end{equation}
for $k \in \mbN_0$; and
\begin{equation}\label{esth617}
|\partial_\s^l \partial_x^k  h_\nlz(x, \s)|\leq C \s^{2-l}
\end{equation}
for $l\leq 2$ and $k\in \mbN_0$.
\end{lem}
\begin{pf}
 \eqref{esthgen} is a direct consequence of \eqref{estphig6} and \eqref{h614}.

Differentiating \eqref{h614} with respect to $\s$ and applying \eqref{estphi611} gives
\begin{equation}\label{esth01}
| \partial_\s^{2k}  h_\nlz(x, \s)| \leq C |x|,\quad \text{and}\quad   |\partial_\s^{2k+1} h_\nlz(x, \s)| \leq C |x| \s.
\end{equation}
Differentiating \eqref{h614} with respect to $x$ and noting that $\phi_\nlz(x, \s) =\hphi_\nlz(x)(1+h_\nlz(x, \s))$,  we  have
\begin{equation}\label{hdalter}
\begin{aligned}
\frac{\partial h_\nlz}{\partial x}(x, \s) = &\frac{\int_0^1  (V_\nl (\theta x) \hphi_\nlz(\theta x) h_\nlz(\theta x,\s) -\s^2 \phi_\nlz(\theta x, \s))  \theta d\theta}{\int_0^1 \hphi_\nlz'(\theta x) d\theta}\\
& - \frac{h_\nlz(x, \s) \int_0^1 \hphi_\nlz''(\theta x) \theta d\theta }{\int_0^1 \hphi_\nlz'(\theta x) d\theta}.
\end{aligned}
\end{equation}
Then \eqref{esth615} follows from \eqref{esth01} and differentiation of \eqref{hdalter}.

It follows from Lemma \ref{lemh} that $|h_\nlz(x, \s)|\leq C\s^2$ for $x\in [-x_0, x_0]$. This implies that $|\partial_x h_\nlz(x, \s)|\leq C \s^2$ by \eqref{hdalter}.  By induction, we can see that $|\partial_x^k h_\nlz(x, \s)|\leq C\s^2$ for $x\in [-x_0, x_0]$. It follows from \eqref{esth615} that $|\partial_\s h_\nlz(x, \s)|\leq C \s$ for $x\in [-x_0, x_0]$. Moreover,  we have
\begin{equation}
\begin{aligned}
\frac{\partial^2 h_\nlz}{\partial x \partial \s}(x, \s) = &\frac{\int_0^1  (V_\nl (\theta x) \hphi_\nlz(\theta x) \partial_\s h_\nlz(\theta x,\s) -\s^2 \partial_\s\phi_\nlz(\theta x, \s))  \theta d\theta}{\int_0^1 \hphi_\nlz'(\theta x) d\theta}\\
& -\frac{\int_0^1 2 \s \phi_\nlz(\theta x, \s)  \theta d\theta + \partial_\s h_\nlz(x, \s) \int_0^1 \hphi_\nlz''(\theta x) \theta d\theta }{\int_0^1 \hphi_\nlz'(\theta x) d\theta}.
\end{aligned}
\end{equation}
Using a similar induction method, we can show that $|\partial_x^k \partial_\s h_\nlz(x, \s)|\leq C\s$.
\end{pf}

For $x\in [x_0, \delta \s^{-1}]$, it is easy to see $h_\nlz(x, \s)$ is the solution of the integral equation
\begin{equation}\label{halter}
\begin{aligned}
&h_\nlz(x, \s) = h_\nlz(x_0, \s) + \frac{\partial h_\nlz}{\partial x}(x_0, \s) \hphi_\nlo(x_0) \hphi_\nlz(x_0)\\
&\quad-  \frac{\partial h_\nlz}{\partial x}(x_0, \s) \hphi_\nlz^2(x_0)\frac{\hphi_\nlo(x)}{\hphi_\nlz(x)}\\
&\quad- \s^2 \int_{x_0}^x \left(\hphi_\nlz(y)\hphi_\nlo(y) -\frac{\hphi_\nlo(x)}{\hphi_\nlz(x)}\hphi_\nlz^2(y)\right)(1+h_\nlz(y,\s)) dy.
\end{aligned}
\end{equation}
For $x\in [-\delta \s^{-1}, -x_0]$, there is a similar representation for $h_\nlz(x, \s)$ as
\begin{equation}\label{halter2}
\begin{aligned}
&h_\nlz(x, \s) = h_\nlz(-x_0, \s) + \frac{\partial h_\nlz}{\partial x}(-x_0, \s) \hphi_\nlo(-x_0) \hphi_\nlz(-x_0)\\
&\quad- \hphi_\nlz^2(-x_0) \frac{\partial h_\nlz}{\partial x}(-x_0, \s)
\frac{\hphi_\nlo(x)}{\hphi_\nlz(x)}\\
&\quad- \s^2 \int_x^{-x_0} \left(\hphi_\nlz(y)\hphi_\nlo(y) -\frac{\hphi_\nlo(x)}{\hphi_\nlz(x)}\hphi_\nlz^2(y)\right)(1+h_\nlz(y,\s)) dy.
\end{aligned}
\end{equation}

Applying this representation of $h_\nlz$, we have the following estimate for $h_\nlz$.
\begin{lem}
 For  $x\in [-\delta \s^{-1}, -x_0]\cup [x_0, \delta \s^{-1}]$,  we have
\begin{equation}\label{refinehest}
| \partial_\s^{2k} h_\nlz(x, \s)| \leq C |x|^{2k},\qquad \text{and}\qquad   |\partial_\s^{2k+1}h_\nlz(x, \s)| \leq C |x|^{2k+2} \s
\end{equation}
 for $k\in \mbN_0$.
\end{lem}
\begin{pf}
We estimate the first three terms on the right hand sides of \eqref{halter} and \eqref{halter2}  by Lemma \ref{lemh63}. The fact that
\begin{equation}
|\hK(x, y, \s)|\leq C|x| \quad \text{for}\quad x\leq y \leq -x_0\quad \text{ or }\quad x_0\leq y\leq x,
\end{equation}
where $\hK$ is defined in  \eqref{defhK},
makes the proof of this lemma similar to that of  Lemma \ref{lem62} .
\end{pf}

Furthermore, we have the following estimates for the derivatives of $h_\nli$.
\begin{lem}\label{lemdeh}
For $x\in [-\delta \s^{-1}, \delta \s^{-1}]$,
\begin{equation}\label{estdehnlz}
|\partial_x^k \partial_{\s}^l h_\nlz(x,\s)|\leq C \lxr^{2-k}\s^{2-l},
\quad \text{and}\quad
|\partial_x^k \partial_{\s}^l h_\nlo(x,\s)|\leq C \lxr^{2-k}\s^{2-l}.
\end{equation}
for $k\in \mbN_0$ and $l\in \mbN_0$.
\end{lem}
\begin{pf}
First, it follows from Lemma \ref{lemh} that \eqref{estdehnlz} holds on $[-\delta \s^{-1}, \delta \s^{-1}]$  in the case $k=l=0$. For $x\in [-x_0, x_0]$, the estimate for $h_\nlz$ in \eqref{estdehnlz} has already been proved in Lemma \ref{lemh63}. Then one can prove  \eqref{estdehnlz} for $x\in [x_0, \delta \s^{-1}]$ by a method similar to \cite[Proposition 4.1]{SofferPrice}. Similarly, we can prove the estimate for $h_\nlz$ for $x\in [-\delta \s^{-1}, -x_0]$.

Let us prove the estimate for $h_\nlo$ for $x\in [-\delta \s^{-1}, -x_0]$. It follows from \eqref{defhnlo} that
\begin{equation}
\begin{aligned}
& h_\nlo(x, \s)=  h_\nlz(x, \s) \\
&\quad+\frac{\hphi_\nlz}{\hphi_\nlo}(x)(1+h_\nlz(x, \s)) \left[  \int_{x_0}^{\delta \s^{-1}} \hphi_\nlz^{-2}(y) \tilde{h}_\nlz(y,\s)dy \right.\\
&\quad \left.+  \int_{-x_0}^{x_0} \frac{\int_0^1 (1-\theta)  \partial_x \tilde{h}(\theta y, \s)d\theta}{(\int_0^1\hphi_\nlz'(\theta y) d\theta)^2}d y  + \int_x^{-x_0} \hphi_\nlz^{-2}(y) \tilde{h}_\nlz(y,\s)dy \right]\\
=&O(x^2 \s^2)+O(x) (1+O(x^2 \s^2))[O(\s^{4\lambda}) +O(\s^2)  +O(\s^2 x)],
\end{aligned}
\end{equation}
where the $O-$terms are of symbol type. Thus  differentiation yields the estimate  \eqref{estdehnlz}. The estimate for $x\in [-x_0, x_0]$ is easier, so we omit it.
\end{pf}

Using the estimates in \eqref{estdehnlz} for $h_\nlz(x, \s)$, we have the following estimate for $h_\nlo(x, \s)$.
\begin{lem}\label{lemfineder}
For all $\s \in [0, \s_1]$ and $x\in [-\delta \s^{-1}, \delta \s^{-1}]$,  $h_\nlo(x, \s)$ satisfies the estimates
\begin{equation}\label{h628}
|\partial_{\s}^{2k} h_\nlo (x, \s) |\leq C_k \lxr^{2k}\quad \text{and} \quad |\partial_{\s}^{2k+1} h_\nlo (x, \s) |\leq C_k \lxr^{2k+2}\s
\end{equation}
for  $k \leq \lambda-1$ and
\begin{equation}\label{h629}
|\partial_{\s}^{2\lambda+l} h_\nlo (x, \s) |\leq C_l \lxr^{2\lambda}\s^{-l}
\end{equation}
for $l \geq 0$.
\end{lem}
\begin{pf}
The proof for the estimates \eqref{h628} and \eqref{h629} on $[x_0, \delta \s^{-1}]$ is similar to that in \cite[Lemma 4.5]{SofferPrice}. Here we give the proof for the estimates \eqref{h628} and \eqref{h629} for $x\in [-\delta \s^{-1}, -x_0]$. For $x\in [-\delta\s^{-1}, -x_0]$, it is easy to see that
\begin{equation}\label{631}
\begin{aligned}
& \frac{\partial^k h_\nlo}{\partial \s^k}(x, \s)= \frac{\partial^k h_\nlz}{\partial \s^k}(x, \s) +\sum_{l=0}^k \frac{\hphi_\nlz}{\hphi_\nlo}(x)\frac{\partial^l}{\partial \s^l}(1+h_\nlz(x, \s)) \times\\
&\quad \left[ \frac{\partial^{k-l}}{\partial \s^{k-l}} \int_{x_0}^{\delta \s^{-1}} \hphi_\nlz^{-2}(y) \tilde{h}_\nlz(y,\s)dy +  \int_{-x_0}^{x_0} \frac{\int_0^1 (1-\theta) \partial_\s^{k-l} \partial_x \tilde{h}(\theta y, \s)d\theta}{(\int_0^1\hphi_\nlz'(\theta y) d\theta)^2}d y \right. \\
&\left. \quad +\frac{\partial^{k-l}}{\partial \s^{k-l}} \int_x^{-x_0} \hphi_\nlz^{-2}(y) \tilde{h}_\nlz(y,\s)dy \right] =\sum_{i=1}^4 I_i
\end{aligned}
\end{equation}
Note that
\begin{equation}\label{632}
\begin{aligned}
&\frac{\partial^{k-l}}{\partial \s^{k-l}} \int_{x_0}^{\delta \s^{-1}} \hphi_\nlz^{-2}(y) \tilde{h}_\nlz(y,\s)dy = \int_{x_0}^{\delta \s^{-1}} \hphi_\nlz^{-2}(y) \frac{\partial^{k-l}\tilde{h}_\nlz}{\partial \s^{k-l}}(y,\s)dy\\
&\quad +\sum_{j=0}^{k-l-1} \frac{d^j}{d\s^j} (\hphi_\nlz^{-2}(\delta \s^{-1}) \partial_\s^{k-l-1-j}\tilde{h}_\nlz(\delta \s^{-1}, \s) (-\delta \s^{-2}))
\end{aligned}
\end{equation}
and 
\begin{equation*}
\left|\int_{x_0}^{\delta \s^{-1}} \hphi_\nlz^{-2}(y) \tilde{h}_\nlz(y,\s)dy \right|\leq C \s^2.
\end{equation*}
If $k =2\lambda$, then it follows from \eqref{refinehest} that
\begin{equation*}
|I_2| \leq C \lxr\left( \lxr^{2\lambda} \s^2 + \sum_{l=0}^{2\lambda-1} \lxr^{l}+ \sum_{l=0}^{2\lambda} \left[\lxr^l \sum_{j=0}^{2\lambda-l-1}\partial_\s^j (\s^{2(\lambda+1)}\s^{l+1+j-2\lambda}\s^{-2})\right] \right) \leq  C\lxr^{2\lambda}.
\end{equation*}
The estimates for the terms $I_1$, $I_3$, and $I_4$ are much easier, so we omit the proofs. This proves \eqref{h629} for $l=0$. The proof for \eqref{h629} with $l\geq 1$ and \eqref{h628} follows from the same strategy combined with \eqref{refinehest}.
\end{pf}

As  in \cite{SofferPrice}, we introduce the following notation.
\begin{defn}\label{def}
For $k\in \mbN_0$ we denote $f\in S_k(x)$ if, for a constant $s>0$,
\begin{enumerate}
\item $f: (0, s) \to \mbR$ is smooth,
\item $|f^{(l)}(x)|\leq  C_l$ for $l\leq k$ and all $x\in (0, s)$,
\item $|f^{(k+l)}(x)|\leq C_l x^{-l}$ for all $l\in \mbN_0$ and $x\in (0, s)$.
\end{enumerate}
We write $f(x)=O_{2k}(1)$ if $f\in S_{2k}(x)$ also satisfies $\lim_{x\to 0+} f^{(2l-1)}(x)=0$ for all $1\leq l\leq k$. Similarly, we write $f(x)=O_{2k+1}(x)$ if $f\in S_{2k+1}(x)$ also satisfies
 $\lim_{x\to 0+} f^{(2l)}(x)=0$ for all $1\leq l\leq k$.
If, in addition, $f(x)\in O_{2k}(1)$ (respectively $O_{2k+1}(x)$) satisfies $f(0)=0$ (respectively $f'(0)=0$), then we write $f(x)\in O_{2k}^0(1)$ (respectively $O_{2k+1}^0(x)$).
\end{defn}

The properties of the functions can be stated via these notations as
\begin{prop}\label{propfun}
 For a given function $f:(0, s)\to \mbR$, one has
\begin{enumerate}
\item If $f\in O_{2k}^0 (1)$, then $\frac{f(x)}{x}\in O_{2k-1}(x)$.
\item If $f\in O_{2k+1}(x)$, then $f'(x)\in O_{2k}(1)$.
\item If $f\in O_{2k}(1)$ and $g\in O_{2l}(1)$ for some $k\ge l$, then $fg\in O_{2l}(1)$.
\item If $f\in O_{2k+1}(\s)$ and $g\in O_{2l}(1)$ for some $k\ge l$, then $fg\in O_{2l+1}(\s)$.
\end{enumerate}
\end{prop}
\begin{pf} (2) and (3) are direct consequences of the definitions. In order to prove (1),  we need only to use the following identity: for $f\in O_{2k}^0(1)$,
\[
\frac{f(x)}{x} =\int_0^1 f'(\theta x) d\theta.
\]

Using Leibniz' rule, we can check easily that $\frac{d^{2i}}{dx^{2i}}(fg)=0$ for $i=0, 1, \cdots, l$. We need only to check that $\frac{d^{2l+1}}{dx^{2l+1}}(fg)$ is bounded. This follows from the estimate
\[
|\frac{d^{2l+1}}{dx^{2l+1}}(fg)|= |\sum_{i=0}^{2l+1}\frac{(2l+1)!}{i!(2l+1-i)!}\frac{d^i f}{dx^i} \frac{d^{2l+1-i}g}{dx^{2l+1-i}}|\leq C |x|\cdot |x|^{-1} +C\leq C.
\]
\end{pf}

\begin{cor}\label{cor6}
 $\phi_\nlz(0, \cdot)=O_{2k}(1)$ and $\partial_x\phi_\nlz(0, \cdot)\in O_{2k}(1)$ for any $k\in \mbN$. $\phi_\nlo(x_0, \cdot)\in O_{2\lambda}(1)$ and $\partial_x\phi_\nlo(x_0, \cdot)\in O_{2\lambda}(1)$  for all $\s\in (0, \s_0)$.
\end{cor}
\begin{pf}
It is easy to see that $\phi_\nlz(0, \s) =\hphi_\nlz(0) (1+h_\nlz(0, \s)) =0$ and
\[
\partial_x\phi_\nlz(0, \s) =\hphi_\nlz'(0) (1+h_\nlz(0, \s))+\hphi_\nlz(0) \partial_x h_\nlz(0, \s) =Q_0.
\]
Thus $\phi_\nlz(0, \s)\in O_{2k}(1)$ and $\partial_x \phi_\nlz(0, \s)\in O_{2k}(1)$ for any $k \in \mbN$.

Since $\phi_\nlo(x_0, \s) =\hphi_\nlo(x_0) (1+h_\nlo(x_0, \s))$, it follows from Lemma \ref{lemfineder} that $\phi_\nlo(x_0, \s) \in O_{2\lambda}(1)$. Moreover,  using $W(\phi_\nlz, \phi_\nlo) =-1$ gives
\[
\partial_x\phi_\nlo(x_0, \s) =\frac{(\hphi_\nlz'(x_0)(1+h_\nlz(x_0, \s))+\hphi_\nlz(x_0)\partial_x h_\nlz(x_0, \s))\phi_\nlo(x_0, \s)-1}{\hphi_\nlz(x_0)(1+h_\nlz(x_0, \s))}.
\]
Therefore, it follows from Lemma \ref{lemh63} that $\partial_x \phi_\nlo(x_0, \s) \in O_{2\lambda}(1)$.
\end{pf}

%\newpage

\subsection{Perturbative Solutions for $\mathbf{|x\s|}$ Large}\label{secperbig}
We construct solutions of equation \eqref{eqmsHmls} in the region $|x\s|$ large. The method is inspired by \cite{SofferPrice}. We first rescale equation (\ref{eqmsHmls}) by introducing a new independent variable $z=\s x$. Upon setting $\tphi(z, \s)=\phi(\s^{-1}z, \s)$,  equation (\ref{eqmsHmls}) is equivalent to
\begin{equation}\label{eqtphi}
\tphi'' +\left(1-\frac{\lambda (\lambda +1)}{z^2}\right)\tphi=\s^{-2} U_{\nl}(\s^{-1}z)\tphi
\end{equation}
where
\begin{equation*}
U_{\nl}(x)=V_{\nl}(x)-\frac{\lambda(\lambda+1)}{x^2}.
\end{equation*}
Note that the differential equation
\begin{equation*}
w'' + \left(1-\frac{\lambda (\lambda +1)}{z^2}\right)w=0
\end{equation*}
is a standard Bessel equation, which
has the fundamental system $\{\sqrt{z}J_{\lambda+1/2}(z), \sqrt{z}Y_{\lambda+1/2}\}$ where $J_{\lambda+1/2}$ and $Y_{\lambda+1/2}$ are Bessel functions (cf. \cite{Askey}). We have the following lemma on the solutions of \eqref{eqtphi}, whose proof is similar to the combination of  \cite[Lemmas 5.1 and  5.2]{SofferPrice}.
\begin{lem}\label{lemvphi}
There exists a smooth solution $\vphi_{\lambda}(z, \s)$ of the equation (\ref{eqtphi}) such that
\begin{equation}\label{ansatzvphi}
\vphi_{\lambda}(z, \s)=\beta_\lambda \sqrt{z}H_{\lambda+1/2}(z)(1+\mfh_{\lambda}(z, \s))
\end{equation}
where $H_{\lambda+1/2}=J_{\lambda+1/2}+iY_{\lambda+1/2}$ is a Hankel function and $\beta_{\lambda}=i\sqrt{\pi/2}e^{i\pi \lambda/2}$. The function $\mfh_{\lambda}(z, \s)$ satisfies
\begin{equation}\label{estb}
|\mfh_{\lambda}(z, \s)|\leq C\langle z\rangle ^{-2+\ep}\s^{1-(2\lambda+3)\ep}
\end{equation}
for all $\s\in (0, \s_2)$ and $z\in [\s^{\ep}, \infty)$ for any $\ep \in (0, \frac{1}{2\lambda +3})$. Furthermore,
\begin{equation*}
|\partial_z^k \partial_\s^l \mfh_\lambda(z, \s)|\leq C_{k, l} z^{-k} \s^{1-(2\lambda+3)\ep-l}
\end{equation*}
for all $z\in [\s^{\ep}, 1]$ and
\begin{equation*}
|\partial_z^k \partial_\s^l \mfh_\lambda(z, \s)|\leq C_{k, l} z^{-2+\ep-k} \s^{1-\ep-l}
\end{equation*}
for all $z\in [1, \infty)$ and any $\ep\in (0, \frac{1}{2\lambda+3})$.
\end{lem}

%\newpage
\section{Representation of the Jost Solutions}\label{secmatch}
\subsection{Representation of $f_\lambda^+$}\label{subsec71}
Note that $\beta_\lambda \sqrt{z}H_{\lambda+1/2}(z) \sim e^{iz}$ (\cite{Askey}), thus the solution $\vphi_\lambda$ constructed in Lemma \ref{lemvphi}  has the asymptotic behavior $\vphi_{\lambda}(z, \s) \sim e^{iz}$ for $z\rightarrow \infty$. This implies
\begin{equation}\label{matchfp}
f^+_\lambda(x, \s)=\vphi_\lambda(\s x, \s).
\end{equation}
Thus we have found a representation of the Jost solution $f^+_\lambda(x, \s)$ which is valid for all $\s\in (0, \s_2)$ and all $x$ with $\s^{-1+\ep}\leq x< \infty$. By appropriately choosing $\ep$, we can always accomplish $\s^{-1+\ep}\leq \delta \s^{-1}$ for any given $\delta>0$ and all $\s\in (0, \s_0)$ for some $\s_0\leq \min\{\s_1, \s_2\}$. Thus, at $x=\s^{-1+\ep}$ for $\s\leq \s_0$, we can represent the solution $f^+_\lambda(\cdot, \s)$ by  $\phi_{\nli}(\cdot, \s)$ ($i=0, 1$) in the following lemma.
\begin{lem}\label{lemclp}
The Wronskians $c_{\li}^+(\s)=W(f_{\lambda}^+(\cdot, \s), \phi_{\nli}(\cdot, \s))$ ($i=0, 1$) for $\s\in (0, \s_0)$ have the asymptotic expansion
\begin{equation}\label{clzp}
\begin{aligned}
c_{\lz}^+(\s)=i\frac{\alpha_\lz\beta_{\lambda}}{B_0}
\sigma^{-\lambda}(1+O(\s^{\ep})+ i O(\s^{(2\lambda+2)\ep})),
\end{aligned}
\end{equation}
and
\begin{equation}\label{clop}
\begin{aligned}
c_{\lo}^+(\s)=- \alpha_\lo\beta_{\lambda}B_0
\sigma^{\lambda+1}(1+O(\s^{\ep})+i O(\s^{-2\lambda \ep})),
\end{aligned}
\end{equation}
for sufficiently small $\ep>0$ where $O-$terms are of symbol type and  $\alpha_\li$ ($i=0, 1$) are nonzero constants  given by
\begin{equation}\label{defalpha}
\alpha_\lo=\frac{1}{\Gamma(\lambda+\frac{3}{2}) 2^{\lambda+\frac{1}{2}}} ,\quad \alpha_\lz=-\frac{\Gamma(\lambda+\frac{1}{2})2^{\lambda+\frac{1}{2}}}{\pi}.
\end{equation}
\end{lem}

The proof of  Lemma \ref{lemclp} is similar to that for \cite[Lemma 6.1]{SofferPrice}. The only difference between these two lemmas is the constant $B_0$ in  \eqref{clzp} and \eqref{clop}, which comes from \eqref{hphipbehavior}.

\subsection{Representation of  $f_\lambda^-(x, \s)$} In Lemma \ref{lemflm}, we constructed $\flm(x, \s)$ for $x\in (-\infty, a)$ for $a\in \mbR$. Note that the linearly independent solutions $\phi_\nli(x, \s)$ ($i=0, 1$) are also constructed on $[-\delta \s^{-1}, \delta \s^{-1}]$. Therefore, we can represent the Jost solution $\flm(x, \s)$ by the linearly independent solutions $\phi_\nli$ ($i=0, 1$).
\begin{lem}\label{lemclm}
The Wronskians $c_{\li}^-(\s)=W(f_{\lambda}^-(\cdot, \s), \phi_{\nli}(\cdot, \s))$ ($i=0, 1$) have the asymptotic behavior
\begin{equation}\label{clzm}
c_{\lz}^-(\s)=O_{2k}(1)+i O_{2k+1}(\s)\quad \text{and}\quad c_{\lo}^-(\s)=O_{2\lambda}^0(1)+i O_{2\lambda +1}(\s)
\end{equation}
for $\s \in (0, \s_0)$ and any $k\in \mbN$.
Furthermore, we have
\begin{equation}\label{clmzero}
c_{\lz}^{-}(0)=1,\quad \text{and}\quad \frac{d c_{\lo}^{-}}{d\s}(0)=i.
\end{equation}
\end{lem}
\begin{pf}
The proof for \eqref{clzm} except for $c_\lo^-(0)=0$ is quite similar to the proof of \cite[Lemma 6.4]{SofferPrice}. It follows from Lemma \ref{lemflm} that $\flm(x, \s)$ is smooth near $\s=0$ and by definition  $\flm(x, \s) =\overline{\flm(x, -\s)}$ for $\s \in \mbR$. In particular, this implies that for fixed $x$ $\Re \flm(x, \s)=O_{2k}(1)$ and $\Im \flm(x, \s) =O_{2k+1}(\s)$ for all $k \in \mbN$. Corollary \ref{cor6} implies  $\phi_\nlz(0, \s) =O_{2k}(1)$ and $\partial_x \phi_\nlz (0, \s) =O_{2k}(1)$. This shows that
\begin{equation*}
c_\lz^{-}=O_{2k}(1)(O_{2k}(1)+i O_{2k +1}(\s))= O_{2k}(1)+i O_{2k +1}(\s)
\end{equation*}
Similarly, Corollary \ref{cor6} and Proposition \ref{propfun}  imply that
\[
c_\lo^-=O_{2\lambda}(1) (O_{2k}(1)+i O_{2k +1}(\s)) = O_{2\lambda}(1)+ i O_{2\lambda  +1}(\s).
\]

We now prove $c_\lo^-(0)=0$ and  \eqref{clmzero}.
Using \eqref{defGlpm} and Lemma \ref{lemh} gives
\begin{equation}\label{compclz}
\begin{aligned}
& c_{\lz}^-(\s)
=  f_\lambda^-(x, \s) \frac{\partial\phi_{\nlz}}{\partial x}(x, \s) - \frac{\partial f_\lambda^-}{\partial x}(x, \s)
\phi_{\nlz}(x, \s)\\
=&  e^{-i \s x} (1+q_\lambda^-(x, \s)) (  \hphi_\nlz'(x) (1+h_{\nlz}(x, \s)) +  \hphi_\nlz(x) \frac{\partial h_{\nlz} } {\partial x}(x, \s) )\\
& -\left[(-i \s)(1+q_\lambda^-(x, \s))+\frac{\partial q_\lambda^-}{\partial x} (x, \s)\right]e^{-i\s x} \hphi_{\nlz}(x) (1+h_{\nlz}(x, \s))\\
=& \sum_{i=1}^2 I_{0i}.
\end{aligned}
\end{equation}
Choose $x= -\s^{-\ep}$. By Lemma \ref{lemhphi}, we have
\begin{equation}\label{match1n}
\begin{aligned}
&\hphi_\nlz(-\s^{-\ep}) =-\s^{-\ep}(1+O(\s^{\ep})), \quad \hphi_\nlz'(-\s^{-\ep}) =1+O(e^{-\frac{\s^{-\ep}}{8M}}).
\end{aligned}
\end{equation}
It follows from Lemmas \ref{lemh}  and  \ref{lemdeh} that
\begin{equation}\label{match2n}
|h_\nlz(-\s^{-\ep}, \s)|\leq C \s^{2- 2\ep},\quad \left|\frac{\partial
h_\nlz}{\partial x}(-\s^{-\ep}, \s)\right|\leq C \s^{2-\ep}.
\end{equation}
Furthermore, Lemma \ref{lemflm} gives
\begin{equation}\label{match3n}
|\qlm(-\s^{-\ep}, \s)|\leq C e^{-\frac{\s^{-\ep}}{8M}},\quad
\left|\frac{\partial \qlm}{\partial x}(-\s^{-\ep}, \s)\right|\leq C
e^{-\frac{\s^{-\ep}}{8M}}.
\end{equation}
Therefore,  combining \eqref{match1n}-\eqref{match3n} gives
\begin{equation*}
\begin{aligned}
I_{01}= & e^{i \s^{1-\ep}} (1+\OC(e^{-\frac{\s^{-\ep}}{8M}})) [(1+ O(e^{-\frac{\s^{-\ep}}{8M}})) (1+O(\s^{2-2\ep}))
-\s^{-\ep}(1+O(\s^{\ep})) O(\s^{2-\ep})]\\
=&1+\OC(\s^{1-\ep}).
\end{aligned}
\end{equation*}
Furthermore,  we have
\begin{equation*}
\begin{aligned}
I_{02} = & -\left [(-i\s)(1+\OC(e^{-\frac{\s^{-\ep}}{8M}})) + \OC(e^{-\frac{\s^{-\ep}}{8M}}) \right] e^{i\s^{1-\ep}} (-\s^{-\ep})(1+O(\s^{\ep}))(1+O(\s^{2-2\ep}))\\
 = &-i\s^{1-\epsilon} (1+\OC(\s^{\ep})).
\end{aligned}
\end{equation*}
Combining the estimates for $I_{01}$ and $I_{02}$ gives  $c_\lz^-(\s)=1+\OC(\s^{1-\ep})$. Thus $c_\lz^-(0)=1$. This finishes the first part of \eqref{clmzero}.

Similarly, we have
\begin{equation}\label{estclo}
\begin{aligned}
&c_{\lo}^-(\s)=  f_\lambda^-(x, \s) \frac{\partial\phi_{\nlo}}{\partial x}(x, \s) - \frac{\partial f_\lambda^-}{\partial x}(x, \s)
\phi_{\nlo}(x, \s)\\
= & e^{-i \s x} (1+q_\lambda^-(x, \s)) \left[ \hphi_\nlo'(x) (1+h_{\nlo}(x, \s))+  \hphi_\nlo(x) \frac{\partial h_{\nlo} } {\partial x}(x, \s)\right]\\
& -\left[(-i \s)(1+q_\lambda^-(x, \s))+\frac{\partial q_\lambda^-}{\partial x} (x, \s)\right]e^{-i\s x} \hphi_{\nlo}(x) (1+h_{\nlo}(x, \s))\\
=& \sum_{i=1}^2 I_{1i}
\end{aligned}
\end{equation}
Choose $x=-\s^{-\ep}$. By Lemma \ref{lemhphi},
\begin{equation}\label{match4n}
\begin{aligned}
&\hphi_\nlo(-\s^{-\ep}) = 1+O(e^{-\frac{\s^{-\ep}}{8M}} ),\quad \text{   and   }\quad \hphi_\nlo'(-\s^{-\ep}) = O(e^{-\frac{\s^{-\ep}}{8M}} ).
\end{aligned}
\end{equation}
It follows from Lemmas \ref{lemh}  and  \ref{lemdeh} that
\begin{equation}\label{match5n}
|h_\nlo(-\s^{-\ep}, \s)|\leq C \s^{2- 2\ep},\quad \left|\frac{\partial
h_\nlo}{\partial x}(-\s^{-\ep}, \s)\right|\leq C \s^{2-\ep}.
\end{equation}
Combining \eqref{match4n}, \eqref{match5n}, and \eqref{match3n} gives
\begin{equation*}
\begin{aligned}
I_{11}=& e^{i\s^{1-\ep}} (1+\OC(e^{-\frac{\s^{-\ep}}{8M}}))
[O(e^{-\frac{\s^{-\ep}}{8M}}) (1+O(\s^{2-2\ep})) +(1+
O(e^{-\frac{\s^{-\ep}}{8M}}))O(\s^{2-\ep}) ]\\
=& \OC(\s^{2-\ep}).
\end{aligned}
\end{equation*}
Moreover,
\begin{equation*}
\begin{aligned}
I_{12} = & -\left[ (-i\s) (1+ \OC (e^{-\frac{\s^{-\ep}}{8M}})) +\OC(e^{-\frac{\s^{-\ep}}{8M}})\right]  e^{-i \s^{1-\ep}}(1+ O(e^{-\frac{\s^{-\ep}}{8M}})) (1+O(\s^{2-2\ep}))\\
= & i\s (1+ \OC(\s^{1-\ep})).
\end{aligned}
\end{equation*}
Therefore, $c_\lo^{-}(0) =0$. Furthermore, differentiating \eqref{estclo} and evaluating the resulting expression at $x=-\s^{-\ep}$ implies that
\[
\frac{d c_\lo^-}{d\s} =i (1+\OC(\s^{1-\ep})).
\]
This finishes the proof of \eqref{clmzero}.
\end{pf}

%\newpage
\section{Asymptotic Behavior of $Z_\nl$}\label{secasymp}
\subsection{Green's Function  near Zero Energy}
The definition of $c_\li^\pm$ implies
\begin{equation}\label{81}
f_\lambda^{\pm} (x, \s) = -c_\lo^{\pm}(\s) \phi_\nlz(x,\s) +c_\lz^{\pm}(\s) \phi_\nlo(x, \s),
\end{equation}
where we used $W(\phi_\nlz, \phi_\nlo)=-1$. Therefore,
\begin{equation*}
W(f_\lambda^-, f_\lambda^+)=c_\lo^-(\s)c_\lz^+(\s) -c_\lz^-(\s)c_\lo^+(\s).
\end{equation*}
Set
\begin{equation*}
A_{jk}(\s) =\Im \left[ \frac{c_{\lambda, j}^-(\s)c_{\lambda, k}^+(\s)}{ c_\lo^-(\s)c_\lz^+(\s) -c_\lz^-(\s)c_\lo^+(\s)}\right].
\end{equation*}
Then we have the following lemma
\begin{lem}\label{lemA}
The functions $A_{jk}$ are of the form
\begin{equation}\label{estAij}
\begin{aligned}
&A_{00}(\s)=-\frac{1}{\s}+S_{2\lambda-1}(\s),\quad A_{01}(\s) =O( \s^{2\lambda-1}),\\
&A_{10}(\s)=O(\s^{2\lambda}),\quad A_{11}(\s) =O( \s^{2\lambda+1}),
\end{aligned}
\end{equation}
where $S_{2\lambda-1}$ is given in Definition \ref{def} and  the $O-$terms are of symbol type.
\end{lem}
\begin{pf}
We prove the lemma by straightforward computation. Note that
\begin{equation*}
\begin{aligned}
& A_{00}(\s) =  \Im \left[ \frac{c_\lz^-(\s) c_\lz^+(\s)}{c_\lo^-(\s)c_\lz^+(\s) -c_\lz^-(\s)c_\lo^+(\s)}\right]
=\Im \left[ \frac{1}{\frac{c_\lo^-}{c_\lz^-}-\frac{c_\lo^+}{c_\lz^+}}\right].
\end{aligned}
\end{equation*}
Using Proposition \ref{propfun} gives
\begin{equation*}
\begin{aligned}
\frac{c_\lo^-}{c_\lz^-}=&\frac{c_\lo^- \overline{c_\lz^-}}{|c_\lz^-|^2}= (i \s +O_{2\lambda}^0(1) +i O_{2\lambda+1}^0(\s))(1+O_{2k}^0(1) +i O_{2k+1}(\s))\\
= &i\s +i O_{2\lambda+1}^0 (\s)+\s^2 S_{2\lambda-1}(\s).
\end{aligned}
\end{equation*}
It follows from Lemma \ref{lemclp} that
\begin{equation*}
\frac{c_\lo^+}{c_\lz^+} = -i \frac{\alpha_{\lo}}{\alpha_\lz} B_0^2\s^{2\lambda+1}( 1+O(\s^{\ep})+i O(\s^{-2\lambda\ep})),
\end{equation*}
and so
\begin{equation*}
A_{00} =-\frac{1}{\s}(1+ \s S_{2\lambda-1}(\s) +O(\s^{2\lambda})) =-\frac{1}{\s} +S_{2\lambda-1}(\s).
\end{equation*}

Next
\begin{equation*}
\begin{aligned}
&A_{01}(\s) =  \Im \left[ \frac{c_\lz^-(\s) c_\lo^+(\s)}{c_\lo^-(\s)c_\lz^+(\s) -c_\lz^-(\s)c_\lo^+(\s)}\right]\\
=&\Im\left[\frac{ (1 +O_{2\lambda}^0(1) +i O_{2\lambda+1}(\s))B_0\alpha_\lo\beta_\lambda \s^{\lambda+1} (1+O(\s^{\ep})+iO(\s^{-2\lambda\ep}))}{-\frac{1}{B_0} \alpha_\lz\beta_\lambda \s^{-\lambda+1}(1+i O(\s^{\ep}))}\right]\\
= & O(\s^{2\lambda-2\lambda\ep}),
\end{aligned}
\end{equation*}
so if $\ep\leq \frac{1}{2\lambda}$, then $A_{01} =O(\s^{2\lambda-1})$.

Furthermore, we have
\begin{equation*}
\begin{aligned}
&A_{10}(\s) =  \Im \left[ \frac{c_\lo^-(\s) c_\lz^+(\s)}{c_\lo^-(\s)c_\lz^+(\s) -c_\lz^-(\s)c_\lo^+(\s)}\right]
=  \Im \left[\frac{1}{ 1-\frac{c_\lz^-(\s)}{c_\lo^-(\s)}\frac{c_\lo^+(\s)}{c_\lz^+(\s)}}\right]\\
=&\Im \left[\frac{1}{1+ \frac{(1+O_{2\lambda}^0(1) +i O_{2\lambda+1}(\s)) \alpha_\lo B_0^2 \s^{2\lambda}(1+O(\s^{\ep})+iO(\s^{-2\lambda\ep}))}{(1 +O_{2\lambda}^0(1) +i O_{2\lambda-1}(\s)) \alpha_\lz (1+O(\s^{\ep})+iO(\s^{(2\lambda+2)\ep}))}}\right]\\
=& \Im \left[1- i \frac{\alpha_\lo}{\alpha_\lz} B_0^2 \s^{2\lambda} (1+O(\s^\ep) +i O(\s^{-2\lambda \ep}))\right]  =  O(\s^{2\lambda}).
\end{aligned}
\end{equation*}
Finally,
\begin{equation*}
\begin{aligned}
&A_{11}(\s) =  \Im \left[ \frac{c_\lo^-(\s) c_\lo^+(\s)}{c_\lo^-(\s)c_\lz^+(\s) -c_\lz^-(\s)c_\lo^+(\s)}\right] = \Im \left[\frac{ \frac{ c_\lo^+(\s)}{c_\lz^+(\s)}}{ 1-\frac{c_\lz^-(\s)}{c_\lo^-(\s)}\frac{c_\lo^+(\s)}{c_\lz^+(\s)}}\right]
\\
=&\Im \left[  \frac{\frac{ \alpha_\lo B_0\beta_\lambda \s^{\lambda+1}(1+O(\s^{\ep})+iO(\s^{-2\lambda\ep}))}{i \frac{1}{B_0} \alpha_\lz \s^{-\lambda}(1+O(\s^{\ep})+iO(\s^{(2\lambda+2)\ep}))}} {1-\OC(\s^{2\lambda+1-2\lambda \ep -1}) } \right]\\
= & O(\s^{2\lambda+1})
\end{aligned}
\end{equation*}
This finishes the proof of the lemma.
\end{pf}

\subsection{Oscillatory Integral Estimates at Small Energies}
For small energy contributions in \eqref{cossol} and \eqref{sinsol},  it is useful to note that, for $\s\in \mbR$, $f_\lambda^\pm (x, -\s) =\overline{f_\lambda^\pm(x, \s)}$ by definition of the Jost solutions. This implies
\begin{equation*}
G_\nl(x, x', -\s) =\overline{G_\nl(x, x', \s)}
\end{equation*}
and hence $\Im [G_\nl(x, x', \s)]$ is an odd function. Thus we have
\begin{equation*}
\int_0^{\infty} \s \cos(t\s) \Im [G_\nl(x, x', \s)] d\s =\frac{1}{2} \int_\mbR \s \cos(t\s) \Im [G_\nl(x, x', \s)] d\s
\end{equation*}
and similarly for the sine integral. Similarly, we can extend $\phi_{\pm  ,\lambda,i}$ and $c_{\li}^\pm(\s)$ to negative $\s$ according to
\begin{equation*}
 \phi_{\pm,\lambda,i}(x,-\s)=\phi_{\pm,\lambda, i}(x,\s), \quad c_\li^\pm(-\s)=\overline{c_\li^\pm(\s)}.
\end{equation*}

Let $\chi\in C_0^{\infty}(\mbR)$ satisfy
\begin{equation}
\chi(x)=\left\{
\begin{aligned}
&1\quad \text{if}\,\,\,\,|x|<1/2,\\
&0\quad \text{if}\,\,\,\, |x|\geq 1,
\end{aligned}
\right.
\end{equation}
and set $\chi_s(x) = \chi\left(\frac{x}{s}\right)$ for any $s>0$.

For $0<\delta<\delta_1$ and $\s_0>0$ as in the first paragraph in \S \ref{subsec71},  define $F(x, x', \s)=  \chi_\delta(\s x)\chi_\delta(\s x')\chi_{\s_0}(\s)$. It is easy to see that $F=0$ if  $ |\s|> \mfs(x, x') := \min\{\delta|x|^{-1}, \delta |x'|^{-1}, \s_0\}$. Furthermore, we have
\begin{equation}\label{estF}
|\partial_3^kF(x, x', \s)|\leq C (\lxr^k + \lxpr^k).
\end{equation}

\begin{lem}\label{lemcsmallen}
Let  $J_1= \int_\mbR  \s \cos(t\s) \Im [G_\nl(x, x', \s)] F(x, x', \s)  d\s$. Then
\begin{equation*}
|J_1| \leq C\lxpr^{2\lambda} \lxr^{2\lambda}  t^{-2\lambda}.
\end{equation*}
\end{lem}
\begin{pf}
By the definition of the Green's function in \eqref{defGreen}, it is easy to see that
\begin{equation*}
\begin{aligned}
J_1 =&  \int_{\mbR} \s \cos(t\s) A_{00}(\s) \phi_{\nlo}(x, \s) \phi_{\nlo}(x', \s) F(x, x', \s) d\s\\
& - \int_{\mbR} \s \cos(t\s) A_{01}(\s) (\phi_{\nlo}(x', \s) \phi_{\nlz}(x, \s)\Theta(x-x') \\
&\qquad + \phi_{\nlo}(x, \s) \phi_{\nlz}(x', \s)\Theta(x'-x))  F(x, x', \s)  dx'd\s\\
& -  \int_{\mbR} \s \cos(t\s) A_{10}(\s)  (\phi_{\nlz}(x', \s) \phi_{\nlo}(x, \s)\Theta(x-x') \\
& \qquad + \phi_{\nlz}(x, \s) \phi_{\nlo}(x', \s)\Theta(x'-x))  F(x, x', \s)d\s\\
& + \int_{\mbR} \s \cos(t\s) A_{11}(\s) \phi_{\nlz}(x, \s) \phi_{\nlz}(x', \s)F(x,x', \s)d\s\\
= & \sum_{i=1}^4 J_{1i}.
\end{aligned}
\end{equation*}

Let $\omega_1(x,x',\s)=  \phi_{\nlo}(x, \s) \phi_{\nlo}(x', \s)  F(x, x', \s)$.
Then it follows from Lemma \ref{lemfineder} and \eqref{estF} that
\[
|\partial_\s^{2\lambda} \omega_1(x, x', \s)|\leq C \lxr^{2\lambda}\lxpr^{2\lambda}.
\]
Integrating by parts $2\lambda$ times gives
\begin{equation*}
\bar{J}_{11} = \int_\mbR \cos(\s t) \omega_1(x, x',\s) d\s =\frac{(-1)^{\lambda}}{t^{2\lambda}}\int_\mbR \cos(\s t) \partial_\s^{2\lambda} \omega_1(x,x',\s) d\s.
\end{equation*}
Therefore,
\[
|\bar{J}_{11}|\leq C t^{-2\lambda} \lxr^{2\lambda}  \lxpr^{2\lambda}.
\]
Since $A_{00}+\frac{1}{\s}\in S_{2\lambda-1}(\s)$, integrating by parts $2\lambda$ times yields
\begin{equation*}
\begin{aligned}
|\tilde{J}_{11}|=& \left|\int_\mbR \s \cos(t\s)(A_{00}+\frac{1}{\s}) \omega_1(x, x',\s) d\s\right|\\
 = &\left|\frac{(-1)^{2\lambda}}{t^{2\lambda}}\int_\mbR \cos( \s t) \partial_\s^{2\lambda}( \s (A_{00}+\frac{1}{\s}) \omega_1(x,x',\s)) d\s\right|\\
 \leq & Ct^{-2\lambda} \lxr^{2\lambda}\lxpr^{2\lambda} .
 \end{aligned}
\end{equation*}
Therefore,
\begin{equation}
|J_{11}|= |\bar{J}_{11}+\tilde{J}_{11} |\leq Ct^{-2\lambda} \lxr^{2\lambda} \lxpr^{2\lambda}.
\end{equation}

In view of $A_{01}(\s) =O(\s^{2\lambda-1})$ and $|\hphi_\nlz|\leq C\lxr^{\lambda+1}$, we have
\begin{equation}
\begin{aligned}
&|\int_\mbR \s \cos(t\s) A_{01}(\s) \phi_\nlo(x', \s) \phi_\nlz(x, \s) F(x,x', \s) d\s| \\
=&t^{-2\lambda}| \int_\mbR \cos(t \s) \partial_\s^{2\lambda}(\s A_{01}(\s) \phi_\nlo(x', \s) \phi_\nlz(x, \s) F(x, x', \s) )d\s|\\
\leq & Ct^{-2\lambda} \left|\int_{-\mfs(x,x')}^{\mfs(x, x')}  \cos(t \s) \hphi_\nlz(x) \sum_{k=0}^{2\lambda} \partial_\s^{k}(\s A_{01}(\s)) \times \right.\\
&\quad  \partial_\s^{2\lambda-k}( \phi_\nlo(x', \s) (1+h_\nlz(x, \s)) F(x, x', \s) )d\s\Bigg|\\
\leq & C t^{-2\lambda} \sum_{k=0}^{2\lambda}\int_{-\mfs(x,x')}^{\mfs(x,x')} \s^{2\lambda-k} \lxr^{\lambda+1} (\lxr^{2\lambda-k}+\lxpr^{2\lambda -k}) d\s\\
\leq & Ct^{-2\lambda} \lxr^{\lambda+1}.
\end{aligned}
\end{equation}
Similarly,
\[
|\int_\mbR \s \cos(t\s) A_{01}(\s) \phi_\nlo(x, \s) \phi_\nlz(x', \s) F(x,x', \s) d\s| \leq  Ct^{-2\lambda}\lxpr^{\lambda+1}.
\]
Thus $|J_{12}|\leq Ct^{-2\lambda}\lxr^{\lambda+1}\lxpr^{\lambda+1 }$.
Similarly, using integration by parts, we have
\begin{equation*}
\sum_{i=3}^4 |J_{1i}|\leq C \lxpr^{2\lambda}\lxr^{2\lambda} t^{-2\lambda}.
\end{equation*}
This proves the lemma.
\end{pf}

%\newpage

\begin{lem}\label{lemssmallen}
Let  $J_2= \int_\mbR  \sin(t\s) \Im [G_\nl(x, x', \s)]  F(x, x', \s) d\s$. Then
\begin{equation*}
\Big|J_2 + \pi \hphi_\nlz(x') \hphi_\nlo(x)\Big|\leq Ct^{-2\lambda}\lxr^{2\lambda+1}\lxpr^{2\lambda+1}.
\end{equation*}
\end{lem}
\begin{pf}
Straightforward computation gives
\begin{equation*}
\begin{aligned}
J_2=&  \int_{\mbR} \sin(t\s) A_{00}(\s) \phi_{\nlo}(x, \s) \phi_{\nlo}(x', \s) F(x, x', \s) d\s\\
& - \int_\mbR  \sin(t\s) A_{01}(\s)  (\phi_{\nlo}(x', \s) \phi_{\nlz}(x, \s)\Theta(x-x') \\
&\qquad + \phi_{\nlo}(x, \s) \phi_{\nlz}(x', \s)\Theta(x'-x)) F(x, x', \s) d\s\\
& -  \int_{\mbR} \sin(t\s) A_{10}(\s) (\phi_{\nlz}(x', \s) \phi_{\nlo}(x, \s)\Theta(x-x') \\
& \qquad + \phi_{\nlz}(x, \s) \phi_{\nlo}(x', \s)\Theta(x'-x)) F(x, x', \s)d\s\\
& + \int_{\mbR} \sin(t\s) A_{11}(\s) \phi_{\nlz}(x, \s) \phi_{\nlz}(x', \s) F(x, x', \s) d\s\\
= & \sum_{j=1}^4 J_{2i}.
\end{aligned}
\end{equation*}

First, let us estimate $J_{22}$.
Since $A_{01}(\s) =O(\s^{2\lambda-1})$, we have
\begin{equation}
\begin{aligned}
&|\int_\mbR  \sin(t\s) A_{01}(\s) \phi_\nlo(x', \s) \phi_\nlz(x, \s) F(x, x', \s) d\s| \\
=&t^{-2\lambda}| \int_\mbR \sin(t \s) \partial_\s^{2\lambda}\big(A_{01}(\s) \phi_\nlo(x', \s) \phi_\nlz(x, \s) F(x, x', \s)\big) d\s|\\
\leq & t^{-2\lambda}| \int_\mbR \sin(t \s) \partial_\s^{2\lambda}A_{01}(\s) \phi_\nlo(x', \s) \phi_\nlz(x, \s) F(x, x', \s) d\s|\\
& +t^{-2\lambda}| \int_\mbR \sin(t \s) \sum_{k=0}^{2\lambda-1}\partial_\s^{k}A_{01}(\s) \partial_\s^{2\lambda-k} (\phi_\nlo(x', \s) \phi_\nlz(x, \s)F(x, x', \s) ) d\s|\\
 = &R_1+R_2.
\end{aligned}
\end{equation}
It follows from Lemma \ref{lemfineder} and \eqref{estF} that
\begin{equation*}
\begin{aligned}
R_2 \leq & Ct^{-2\lambda} \left|\int_{-\mfs(x,x')}^{\mfs(x, x')}  \cos(t \s) \hphi_\nlz(x) \sum_{k=0}^{2\lambda-1} \partial_\s^{k} A_{01}(\s) \times \right.\\
&\quad  \partial_\s^{2\lambda-k}( \phi_\nlo(x', \s) (1+h_\nlz(x, \s)) F(x, x', \s) )d\s\Bigg|\\
\leq & C t^{-2\lambda} \sum_{k=0}^{2\lambda-1}\int_{-\mfs(x,x')}^{\mfs(x,x')} \s^{2\lambda-1-k} \lxr^{\lambda+1} (\lxr^{2\lambda-k}+\lxpr^{2\lambda -k}) d\s\\
\leq & Ct^{-2\lambda} \lxr^{\lambda+2}\lxpr.
\end{aligned}
\end{equation*}
Let $\omega_2(x, x', \s) =  \partial_\s^{2\lambda}A_{01}(\s) \phi_\nlo(x', \s) \phi_\nlz(x, \s) F(x, x', \s)$. Then $R_1$ can be estimated as
\begin{equation*}
\begin{aligned}
R_1\leq  &|t^{-2\lambda}\int_\mbR \sin(\s t) \omega_2(x, x', \s) \chi(\s t)d\s |+|t^{-2\lambda} \int_\mbR \sin(\s t) \omega_2(x, x', \s) (1-\chi(\s t))d\s|\\
=& R_{11}+R_{12}
\end{aligned}
\end{equation*}
Note that $\omega_2(x, x', \s)$ can be estimated as
\begin{equation}
|\omega_2(x, x', \s) |\leq C\frac{1}{\s} \lxr^{\lambda+1}  \quad\text{and}\quad |\partial_\s \omega_2(x, x', \s) |\leq C \frac{1}{\s^2} \lxr^{\lambda+1}.
\end{equation}
Therefore,
\begin{equation*}
\begin{aligned}
R_{11} \leq & C t^{-2\lambda} \int_\mbR \frac{|\sin (t\s)|}{|\s|}  \lxr^{\lambda+1} \chi(\s t) d\s  \leq  C t^{-2\lambda} \int_\mbR \frac{|\sin \s|}{|\s|}  \lxr^{\lambda+1} \chi(\s ) d\s  \\
 \leq &C t^{-2\lambda} \lxr^{\lambda+1},
 \end{aligned}
 \end{equation*}
 and integrating by parts yields
 \begin{equation*}
 \begin{aligned}
 R_{12}\leq  & C t^{-2\lambda-1} | \int_\mbR \cos(\s t) \partial_\s \omega_2(x, x', \s) (1-\chi(\s t)) d\s |\\
  &+C t^{-2\lambda-1} | \int_\mbR \cos( \s t) \omega_2(x, x', \s) \chi'(\s t) t d\s|\\
\leq & C t^{-2\lambda-1} \int_\mbR \frac{|\cos (\s t)|}{\s^2}\lxr^{\lambda+1} (1-\chi(\s t)) d\s\\
 & +C t^{-2\lambda-1}\int_\mbR \frac{|\cos (\s t)|}{\s} \lxr^{\lambda+1}|\chi'(\s t)|  t d\s\\
\leq & C t^{-2\lambda}\int_{1/2}^\infty  \frac{|\cos \s|}{\s^2}  \lxr^{\lambda+1} d\s  + C t^{-2\lambda} \int_{1/2}^1 \frac{|\cos \s|}{\s} \lxr^{\lambda+1}d\s\\
\leq& Ct^{-2\lambda}  \lxr^{\lambda+1}.
\end{aligned}
\end{equation*}
Thus
\begin{equation*}
|\int_\mbR  \sin(t\s) A_{01}(\s) \phi_\nlo(x', \s) \phi_\nlz(x, \s) F(x, x', \s) d\s| \leq Ct^{-2\lambda}  \lxr^{\lambda+2}\lxpr.
\end{equation*}
Note that $F(x, x', \s) =F(x', x, \s)$,  so
\begin{equation*}
|\int_\mbR  \sin(t\s) A_{01}(\s) \phi_\nlo(x, \s) \phi_\nlz(x', \s) F(x, x', \s) d\s| \leq Ct^{-2\lambda}  \lxpr^{\lambda+2}\lxr.
\end{equation*}
Hence
\begin{equation*}
|J_{22}|\leq Ct^{-2\lambda}  \lxr^{\lambda+2}\lxpr^{\lambda+2}.
\end{equation*}
Using the same method, we have
\begin{equation*}
\sum_{i=3}^4 |J_{2i}|\leq C
 t^{-2\lambda}  \lxr^{2\lambda}\lxpr^{2\lambda}.
\end{equation*}

Now we  analyze  $J_{21}$.  Note that
\begin{equation*}
\begin{aligned}
J_{21} = & \int_{\mbR} \sin (\s t) A_{00}(\s) \phi_{\nlo}(x, \s)\phi_{\nlo}(x', \s)F(x,x', \s)d\s\\
= & - \int_{\mbR} \sin (\s t) \frac{1}{\s} \phi_{\nlo}(x, \s) \phi_{\nlo}(x', \s) F(x, x', \s) d\s\\
&+ \int_{\mbR} \sin (\s t) (A_{00}(\s)+\frac{1}{\s}) \phi_{\nlo}(x,\s)  \phi_{\nlo}(x', \s ) F(x, x', \s)d\s\\
= & \bar{J}_{21}+ \tilde{J}_{21}.
\end{aligned}
\end{equation*}
Using the analysis similar to that for $\tilde{J}_{12}$ and  $J_{22}$, we have
\[
 |\tilde{J}_{21}|\leq Ct^{-2\lambda}\lxr^{2 \lambda}\lxpr^{2 \lambda}.
 \]
Let
\begin{equation*}
\begin{aligned}
E_{ij} =&  \int_{\mbR} \frac{\sin (\s t)}{ \s} \hphi_{\nlo}(x)g_i(x, \s) \hphi_{\nlo}(x') g_j(x', \s) F(x, x', \s) d\s
\end{aligned}
\end{equation*}
where
\begin{equation*}
g_i(x, \s) =\left\{
\begin{aligned}
&1,\,\,\,\, \,\,\qquad \qquad \text{if}\,\, i=0,\\
&h_\nlo(x, \s), \,\, \,\, \text{if}\,\, i=1.
\end{aligned}
\right.
\end{equation*}
Then $\bar{J}_{21}=-\sum_{i,j=0}^1 E_{ij}$.

Since $\int_0^{\infty} \frac{\sin x}{x}dx =\frac{\pi}{2}$ and $F(x, x', 0)=0$, we have
\begin{equation}\label{eqE00}
\begin{aligned}
&E_{00}-\pi \hphi_\nlo(x)\hphi_\nlo(x') =   \int_\mbR \frac{\sin (\s t)}{ \s}  \hphi_\nlo(x) \hphi_\nlo(x') (F(x, x', \s)- 1) d \s\\
=&  \int_\mbR  \sin (\s t)  \hphi_\nlo(x) \hphi_\nlo(x') \int_0^1 \partial_3 F(x, x', \theta \s)d\theta  d\s.
\end{aligned}
\end{equation}
 If $|\s|\geq 1$, then it follows from the definition of $F(x, x', \s)$ that
\begin{equation}\label{estF2}
|\int_0^1 \partial_\s^k F(x, x', \theta \s) d\theta| = |\int_0^{\s_0/|\s|} \partial_\s^k F(x, x', \theta \s) d\theta|\leq C\s^{-1} (\lxr^k + \lxpr^k).
\end{equation}
Integrating by parts $k$ times for \eqref{eqE00} gives
\begin{equation*}
|E_{00}-\pi \hphi_\nlo(x)\hphi_\nlo(x')|\leq Ct^{-k} (\lxr^{k+1} + \lxpr^{k+1})
\end{equation*}
for any $k\in \mbN_0$, where the boundary terms vanish due to \eqref{estF2}.

We are now in a position to estimate $E_{01}$.  Note that $h_\nlo(x, 0)=0$, so
\begin{equation*}
\frac{h_\nlo(x_1, x_2)}{x_2} =\int_0^1 \partial_2 h_\nlo(x_1, \theta x_2) d\theta.
\end{equation*}
Thus
\begin{equation*}
\begin{aligned}
E_{01} =&  \int_{\mbR} \sin (\s t) \hphi_{\nlo}(x) \hphi_{\nlo}(x') F(x, x', \s)\int_0^1 \partial_2 h_\nlo(x', \theta \s) d\theta d\s.
\end{aligned}
\end{equation*}

Let $\omega_3(x, x', \s) =F(x, x', \s) \int_0^1 \partial_2 h_\nlo(x', \theta \s) d\theta$. Obviously, for $|\s|\geq \s_0\geq \mfs(x, x')$, $F(x, x', \s) =0$, thus  $\partial_\s^k \omega_3(x, x',\s) =0$ for $k \in \mbN_0$ and $x$, $x'\in \mbR$. Therefore,
\begin{equation*}
\begin{aligned}
&\int_\mbR \sin(\s t) \omega_3(x, x', \s) d\s  =  t^{-2\lambda} \int_\mbR \sin(\s t) \partial_\s^{2\lambda}\omega_3(x, x', \s) d\s\\
= & t^{-2\lambda} \int_\mbR \sin(\s t) F(x, x', \s) \int_0^1 \partial_2^{2\lambda+1} h_\nlo(x', \theta \s) \theta^{2\lambda} d\theta \chi(\s t) d\s\\
& + t^{-2\lambda} \int_\mbR \sin(\s t) F(x, x', \s) [\int_0^1 \partial_2^{2\lambda+1} h_\nlo(x', \theta \s) \theta^{2\lambda} d\theta ](1-\chi(\s t)) d\s\\
& + t^{-2\lambda}\int_\mbR \sin (\s t)\sum_{k=0}^{2\lambda-1} \partial_3^{2\lambda -k} F(x, x', \s) \int_0^1 \partial_2^{k+1} h_\nlo(x', \theta \s) \theta^{k} d\theta d\s\\
= &\sum_{i=1}^3 S_i.
\end{aligned}
\end{equation*}
First,  using Lemma \ref{lemfineder} and \eqref{estF} gives
\begin{equation*}
\begin{aligned}
|S_3|\leq & Ct^{-2\lambda}\int_{-\mfs(x, x')}^{\mfs(x,x')}\sum_{k=0}^{2\lambda-1} (\lxr^{2\lambda-k}+\lxpr^{2\lambda-k}) \lxpr^{k+1}d \s\\
 \leq &Ct^{-2\lambda} \lxr^{2\lambda+1}\lxpr^{2\lambda+1}.
\end{aligned}
\end{equation*}
It follows from \eqref{h629} that
\begin{equation*}
|\partial_2^{2\lambda+1}h_\nlo(x', \theta \s)|\leq C \frac{1}{|\theta \s |}\lxpr^{2\lambda}.
\end{equation*}
Then
\begin{equation*}
\begin{aligned}
|S_1|\leq &\int_\mbR  t^{-2\lambda}  \lxpr^{2\lambda}\frac{|\sin (\s t)|}{|\s|} \chi(\s t) d\s\leq  t^{-2\lambda}  \lxpr^{2\lambda} \int_\mbR  \frac{|\sin \s |}{|\s|} \chi(\s ) d\s \\
\leq & C  t^{-2\lambda}  \lxpr^{2\lambda}.
\end{aligned}
\end{equation*}
Since $F(x, x', \s) (1-\chi(\s t))=0$ for $|\s|\leq 1/(2t)$ or $|\s|>\s_0$, integrating by parts yeilds
\begin{equation*}
\begin{aligned}
|S_2| \leq &t^{-2\lambda-1} |\int_\mbR \cos (\s t) \partial_\s F(x, x', \s) [\int_0^1 \partial_2^{2\lambda+1} h_\nlo(x', \theta \s) \theta^{2\lambda} d\theta ](1-\chi(\s t)) d\s| \\
&+ t^{-2\lambda-1} |\int_\mbR \cos (\s t)  F(x, x', \s) [\int_0^1 \partial_2^{2\lambda+2} h_\nlo(x', \theta \s) \theta^{2\lambda+1} d\theta ](1-\chi(\s t)) d\s| \\
& +t^{-2\lambda-1} |\int_\mbR \cos (\s t)  F(x, x', \s) [\int_0^1 \partial_2^{2\lambda+1} h_\nlo(x', \theta \s) \theta^{2\lambda} d\theta ]\chi'(\s t) t d\s| \\
= & \sum_{i=1}^3 S_{2i}.
\end{aligned}
\end{equation*}
Now
\begin{equation*}
\begin{aligned}
|S_{21}|\leq & Ct^{-2\lambda-1} \int_{t/2}^{\s_0} \frac{|\cos(\s t)|}{\s} (\lxr + \lxpr) \lxpr^{2\lambda}(1-\chi(\s t)) d\s \\
= &C t^{-2\lambda-1} \int_{1/2}^{\s_0 t}  \frac{|\cos\s|}{\s}(\lxr+\lxpr) \lxpr^{2\lambda}(1-\chi(\s)) d\s \\
 \leq &  Ct^{-2\lambda-1} \ln t (\lxr +\lxpr) \lxpr^{2\lambda} \leq C t^{-2\lambda} \lxr^{2\lambda+1}\lxpr^{2\lambda+1}.
\end{aligned}
\end{equation*}
Next,
\begin{equation*}
\begin{aligned}
S_{22} \leq & t^{-2\lambda-1} \int_\mbR \frac{|\cos(\s t)|}{t} \frac{1}{\s^2} \lxpr^{2\lambda} (1-\chi(\s t))d\s \\
\leq &Ct^{-2\lambda} \lxpr^{2\lambda}\int_{1/2}^\infty \frac{1}{\s^2}d\s \leq Ct^{-2\lambda} \lxpr^{2\lambda}.
\end{aligned}
\end{equation*}
Finally,
\begin{equation*}
\begin{aligned}
S_{23} \leq & t^{-2\lambda-1}\lxpr^{2\lambda} \int_\mbR \frac{|\cos (\s t)|}{\s} |\chi'(\s t)| t d\s\\
\leq & C t^{-2\lambda} \lxpr^{2\lambda} \int_{1/2}^1 \frac{|\cos (\s )|}{\s} |\chi'(\s)| d\s \leq  C t^{-2\lambda} \lxpr^{2\lambda}.
\end{aligned}
\end{equation*}
Therefore
\[
|S_2| \leq C t^{-2\lambda} \lxpr^{2\lambda+1}\lxpr^{2\lambda +1}.
\]
Summing the estimates for $S_i$ (i=1, 2, 3) together, we have
\[
|E_{01}| =|\int_\mbR \sin(\s t) \hphi_\nlo(x)\hphi_\nlo(x')\omega_3(x, x', \s)d\s |\leq C t^{-2\lambda} \lxr^{2\lambda +1}\lxpr^{2\lambda+1}
\]
Similarly, we can prove that
\[
|E_{10}|+|E_{11}|\leq  C t^{-2\lambda} \lxr^{2\lambda+1}\lxpr^{2\lambda+1}
\]
Combining all estimates together finishes the proof of Lemma \ref{lemssmallen}.
\end{pf}

%\newpage

In order to deal with the Green's function that involves $f_\lambda^-(x', \s)$ for $x'\geq 0$ and $f_\lambda^+(x, \s)$ for $x\leq 0$ we have to estimate the reflection and transmission coefficients $a(\s)$ and $b(\s)$ defined in (\ref{trancoe1}). It follows from \eqref{tranbW} that
\begin{equation}\label{best}
\frac{b(\s)}{W(\flm, \flp)(\s)} =-\frac{i}{2\s}.
\end{equation}
The following proposition provides the estimate for $a(\s)$.
\begin{prop}
For $|\s|\leq \s_0$, we have
\begin{equation}\label{smallsaW}
\frac{a(\s)}{W(f_\lambda^-,
f_\lambda^+)(\s)}=-\frac{i}{2\s}(1+\OC(\s^{\ep})),
\end{equation}
where the $\OC-$term is of symbol type.
\end{prop}
\begin{pf}
It follows from (\ref{trancoe1}) that
\begin{equation*}
W(f_\lambda^-, \overline{f_\lambda^+})(\s)=-2i\s a(\s).
\end{equation*}
Therefore,
\begin{equation}\label{smallsaW1}
\frac{a(\s)}{W(f_\lambda^-, f_\lambda^+)(\s)}=-\frac{W(f_\lambda^-,
\overline{f_\lambda^+})}{2i\s W(f_\lambda^-, f_\lambda^+)}.
\end{equation}
From Lemmas \ref{lemclp} and \ref{lemclm} we conclude that
\begin{equation}\label{Wronest}
W(f_\lambda^-, f_\lambda^+)(\s)=
c_\lo^-(\s)c_\lz^+(\s)-c_\lz^-(\s) c_\lo^+(\s)=
-\frac{\alpha_\lz \beta_\lambda}{B_0}  \s^{-\lambda+1} (1+\OC(\s^{\ep}))
\end{equation}
and
\begin{equation*}
W(f_\lambda^-, \overline{f_\lambda^+})(\s)=
c_\lo^-(\s)\overline{c_\lz^+(\s)}-c_\lz^-(\s)\overline{c_\lo^+(\s)}=
\frac{\alpha_\lz \beta_\lambda}{B_0}  \s^{-\lambda+1} (1+\OC(\s^{\ep})).
\end{equation*}
Thus
\begin{equation}\label{smallsaWW}
\frac{W(f_\lambda^-, \overline{f_\lambda^+})(\s)}{W(f_\lambda^-,
f_\lambda^+)(\s)}=-1+\OC(\s^{\ep}).
\end{equation}
Then  estimate (\ref{smallsaW}) is a consequence of (\ref{smallsaW1}) and (\ref{smallsaWW}).
\end{pf}

\begin{lem}\label{lemsxlxp}
Let $F_2(x, x', \s) = \chi_{\s_0}(\s)\chi_\delta(\s x)(1-\chi_\delta(\s x')) $ and  $l\geq 1$. Then we have
\begin{equation}\label{Scosest2}
\begin{aligned}
& \left|\int_\mbR \s \cos(t\s)
\frac{\flm(x',\s)\flp(x,\s)}{W(f_\lambda^-,f_\lambda^+)} F_2(x, x', \s) d\s\right|
\leq  C\ltr^{-l}\lxr^{l+\lambda} \lxpr^{l+\lambda}
\end{aligned}
\end{equation}
and
\begin{equation}\label{Ssinest2}
\begin{aligned}
 \left|\int_\mbR  \sin(t\s)
\frac{\flm(x',\s)\flp(x,\s)}{W(f_\lambda^-,f_\lambda^+)} F_2(x, x', \s) d\s\right|
\leq  C\ltr^{-l+1} \lxr^{l+\lambda} \lxpr^{l+\lambda}.
\end{aligned}
\end{equation}
\end{lem}
\begin{pf}
First, if $x'=0$, then $F_2(x, x', \s)=0$. Thus estimates \eqref{Scosest2} and \eqref{Ssinest2} are trivial.

Second, let $x\in \mbR$, $x'< 0$, $|\s|<\s_0$, $|\s x|\leq \delta$ and $|\s x'|\geq\frac{\delta}{2}$.
Obviously, we have
\begin{equation*}
|\partial_\s^k F_2(x, x', \s) |\leq C(\lxr^k +\lxpr^k).
\end{equation*}
For $\Glp$ in (\ref{defGlpm}),  it follows from \eqref{81} that
\begin{equation*}
\Glp(x, \s)=e^{-i\s x} (-c_\lo^+(\s) \phi_\nlz(x,\s)+c_\lz^+(\s)\phi_\nlo(x,\s)).
\end{equation*}
It follows from Lemmas \ref{lemdeh} and  \ref{lemclp} that
\begin{equation*}
\begin{aligned}
|\partial_\s^k (c_\lo^+(\s)\phi_\nlz(x,\s))| \leq
C \lxr^{\lambda +1} |\s|^{\lambda-k} \leq C  |\s|^{-k}\lxr
\end{aligned}
\end{equation*}
and
\begin{equation*}
\begin{aligned}
|\partial_\s^k (c_\lz^+(\s)\phi_\nlo(x,\s))| \leq   C|\s|^{-\lambda-k},
\end{aligned}
\end{equation*}
which implies
\begin{equation}\label{Glpxs}
|\partial_\s^k \Glp(x,\s)|\leq C|\s|^{-\lambda-k}\quad
\text{for}\,\, |x|\leq \delta \s^{-1}\,\, \text{and}\,\, k\in
\mbN.
\end{equation}
Using Lemma \ref{lemflm}, we have
\begin{equation}\label{Glmxl}
|\partial_\s^k \Glm(x', \s)|\le C_k \quad \text{for  }x'\leq -\frac{\delta}{2} |\s^{-1}|.
\end{equation}

Define
\begin{equation}\label{defomega6}
\omega_4(x, x', \s)=\s
\frac{\Gamma_\lambda^-(x',\s)\Gamma_\lambda^+(x,
\s)}{W(f_\lambda^-,f_\lambda^+)}F_2(x, x', \s).
\end{equation}
Then using $|\s|^{-1}\leq \lxpr$, \eqref{Wronest}, (\ref{Glpxs}), and \eqref{Glmxl} yields
\begin{equation*}
|\partial_\s^k\omega_4(x,x',\s)|\leq C\lxpr^{k}.
\end{equation*}
If $|t+x-x'|\geq \frac{1}{2}t$, we integrate by parts $l$-times
to obtain
\begin{equation}\label{argument1}
\begin{aligned}
&\left|\int_\mbR e^{i\s(\pm
t+x-x')}\omega_4(x,x',\s) d\s\right| \leq C |\pm t+x-x'|^{-l} \lxr^{l}\lxpr^{l} \leq C \ltr^{-l} \lxr^{l}\lxpr^{l},
\end{aligned}
\end{equation}
where we used the fact $\omega_4(x, x', \s)=0$ for $|\s|\leq \frac{\delta}{2|x'|}$.
If $|\pm t+x-x'|\leq \frac{1}{2} t$, we have $\lxr^{-l}
\lxpr^{-l} \leq C\ltr^{-l}$ as $t\rightarrow \infty$ and
thus
\begin{equation}\label{argument2}
\begin{aligned}
&\left|\int_\mbR e^{i\s(\pm
t+x-x')}\omega_4(x,x',\s) d\s\right| \leq C
\ltr^{-l} \lxr^{l}\lxpr^{l}.
\end{aligned}
\end{equation}
Therefore,  we obtain \eqref{Scosest2} for  $x'\leq 0$.

Finally, let $x\in \mbR$, $x' > 0$, $|\s|\leq \s_0$, $|\s x|\leq \delta$ and
$|\s x'|\geq \frac{\delta}{2}$. Using \eqref{trancoe1} and \eqref{81}, we  can represent $\Glm(x',\s)$ as
\begin{equation}
\Glm(x', \s) =a(\s) e^{2i \s x'} \Glp(x', \s) +b(\s)\overline{\Glp(x', \s)}.
\end{equation}
Then $\omega_4(x, x',
\s)$ defined in \eqref{defomega6} becomes
\begin{equation}
\begin{aligned}
\omega_4(x, x', \s)=& \s \frac{(a(\s) e^{2i \s
x'}\Gamma_\lambda^+(x',\s) +b(\s)
\overline{\Gamma_\lambda^+(x',\s)})\Gamma_\lambda^+(x,
\s)}{W(f_\lambda^-,f_\lambda^+)}F_2(x, x', \s).
\end{aligned}
\end{equation}
Furthermore, since
$|\s|$ is small, it follows from Lemma \ref{lemvphi} that
\begin{equation*}
\Gamma_\lambda^+(x', \s) =e^{-i\s x'}\varphi_\lambda(\s x', \s) = e^{-i \s x'} \beta_\lambda \sqrt{\s x'}H_{\lambda+1/2}(\s x')  (1+\mfh_\lambda(\s x', \s))
\end{equation*}
and
\begin{equation}
|\partial_\s^k \mfh_\lambda(\s x',\s)|\leq C\langle x'\rangle^k.
\end{equation}
This implies that
\begin{equation}\label{Glpxl}
|\partial_\s^k \Gamma_\lambda^+(x', \s)|\leq C\lxpr^k\quad \text{for}
\,\, x'>\frac{\delta}{2} \s^{-1}\,\, \text{and for all} \,\, k\in
\mbN.
\end{equation}
The estimates (\ref{best}), (\ref{smallsaW}), \eqref{Glpxs}, and \eqref{Glpxl} yield
\begin{equation}
|\partial_\s^k \omega_4(x, x', \s)|\leq C \lxpr^{k+\lambda}.
\end{equation}
Therefore, as the proof for (\ref{argument1}) and
(\ref{argument2}) for $x'<0$, we get \eqref{Scosest2} for $x'>0$.

For the sine evolution, note that there will be no $\s$ factor in $\omega_4$, therefore, we can integrate by part $l-1$ times, which gives the decay $t^{-l+1}$ in \eqref{Ssinest2}.

\end{pf}

Similarly, we have the following estimate, whose  proof is similar to that for Lemma \ref{lemsxlxp} or that  in \cite[Lemmas 8.3, 8.6, 8.7]{SofferPrice}.
\begin{lem}\label{lemllen}
Let $l\in \mbN_0$. Then the following estimates hold:
\begin{equation*}
\begin{aligned}
&\left|\int_\mbR \s \cos(t\s)
\frac{\flm(x',\s)\flp(x,\s)}{W(f_\lambda^-,f_\lambda^+)} \chi_{\s_0}(\s)(1-\chi_\delta(x\s)) d\s\right|
\leq  C\ltr^{-l}\lxr^{l} \lxpr^{l}
\end{aligned}
\end{equation*}
and
\begin{equation*}\label{Ssinest1}
\begin{aligned}
&\left|\int_\mbR  \sin(t\s)
\frac{\flm(x',\s)\flp(x,\s)}{W(f_\lambda^-,f_\lambda^+)} \chi_{\s_0}(\s) (1-\chi_\delta(x\s))d\s\right|
\leq  C\ltr^{-l+1} \lxr^{l} \lxpr^{l}.
\end{aligned}
\end{equation*}
\end{lem}

%\newpage

\subsection{Oscillatory Integral Estimates for Large Energies}
In the case of $\s\geq \s_0$, there will be no singularity in Green's function $G_{\nl}$ by Lemma \ref{lemwron}.
Using Lemma \ref{lembigs},
similar to \cite[Proposition 9.1 and Corollary 9.2]{SofferPrice},   we have the following two estimates.
\begin{prop}\label{propcl}
Let $l \in \mbN$ and $\s_0>0$ sufficiently small. Then we have the estimates
\begin{equation*}
\begin{aligned}
&\sup_{x\in\mbR} \left|\int_\mbR\int_\mbR \s e^{\pm i t\s} G_\nl(x, x', \s) (1-\chi_{\s_0}(\s))\langle x\rangle^{-l} \langle x'\rangle^{-l}w(x') dx' d\s\right|\\
\leq & C \langle t\rangle^{-l}\int_{\mbR}(|w'(x')|+|w(x')|)dx'
\end{aligned}
\end{equation*}
and
\begin{equation*}
\begin{aligned}
&\sup_{x\in\mbR} \left|\int_\mbR\int_\mbR  e^{\pm i t\s} G_\nl(x, x', \s) (1-\chi_{\s_0}(\s))\langle x\rangle^{-\alpha} \langle x'\rangle^{-\alpha}w(x') dx' d\s\right|\\
\leq & C \langle t\rangle^{-\alpha}\int_{\mbR}|w(x')| dx'
\end{aligned}
\end{equation*}
for all $t\geq 0$ and any $w\in \msS(\mbR)$.
\end{prop}

\subsection{Proof of Theorem \ref{thmZm}}
Note that
\begin{equation*}
\begin{aligned}
& \frac{2}{\pi} \int_0^\infty \int_\mbR \s \cos(t \s)\Im[G_\nl(x, x', \s) ]u_\nl(x') dx' d\s \\
= & \frac{1}{\pi} \int_\mbR \int_\mbR \s \cos(t \s)\Im[G_\nl(x, x', \s) ]u_\nl(x') dx' d\s\\
= & \frac{1}{\pi} \int_\mbR \int_\mbR \s \cos(t \s)\Im[G_\nl(x, x', \s) ]u_\nl(x')  \Big[F(x, x', \s) +F_2(x, x', \s) \\
 &\quad  +\chi_{\s_0}(\s) (1- \chi_\delta(\s x) )    + (1-\chi_{\s_0}(\s)) \Big] dx' d\s\\
= &\sum_{i=1}^4 I_i.
\end{aligned}
\end{equation*}
Lemma \ref{lemcsmallen} implies that
\begin{equation*}
 |I_1(t, x)|\leq C t^{-2\lambda} \lxr^{2\lambda} \|\ldr^{2\lambda } u_\nl(\cdot)\|_{L^1} .
\end{equation*}
Furthermore, Lemma \ref{lemsxlxp}  gives
\begin{equation*}
 |I_2| \leq C t^{-2\lambda} \lxr^{3\lambda}  \|\ldr^{3\lambda} u_\nl(\cdot) \|_{L^1} .
\end{equation*}
It follows from Lemma \ref{lemllen}  and  Proposition \ref{propcl}  that
\begin{equation*}
|I_3|+ |I_4|\leq C t^{-2\lambda} \lxr^{2\lambda} (\| \ldr^{2\lambda} u_\nl(\cdot)  \|_{L^1} +\|\ldr^{2\lambda} u_\nl'(\cdot) \|_{L^1} ).
\end{equation*}
Combining the estimates for $I_i$ ($i=1, \cdots, 4$), we have
\begin{equation}\label{cosest}
\begin{aligned}
&\lxr^{-3\lambda}\left| \frac{2}{\pi} \int_0^\infty \int_\mbR \s \cos(t \s)\Im[G_\nl(x, x', \s) ]u_\nl(x') dx' d\s \right|\\
\leq  &C t^{-2\lambda}\Big( \|  \ldr^{3\lambda} u_\nl(\cdot)\|_{L^1} +\| \ldr^{3 \lambda} u_\nl'(\cdot)\|_{L^1}\Big).
\end{aligned}
\end{equation}

Similar to the cosine integral, we have
\begin{equation*}
\begin{aligned}
& \frac{2}{\pi} \int_0^\infty \int_\mbR  \sin(t \s)\Im[G_\nl(x, x', \s) ]v_\nl(x') dx' d\s\\
=& \frac{1}{\pi} \int_\mbR \int_\mbR  \sin(t \s)\Im[G_\nl(x, x', \s) ]v_\nl(x') dx' d\s\\
= & \frac{1}{\pi} \int_\mbR \int_\mbR  \sin(t \s)\Im[G_\nl(x, x', \s) ]  v_\nl(x') \Big[F(x, x', \s) +F_2(x, x', \s) \\
 &\quad  +\chi_{\s_0}(\s) (1- \chi_\delta(\s x) )    + (1-\chi_{\s_0}(\s)) \Big] dx' d\s\\
= &\sum_{i=1}^4 S_i.
\end{aligned}
\end{equation*}
Lemma \ref{lemssmallen}  implies that
\begin{equation*}
\left |S_1  +D  \hphi_\nlo(x) \right |\leq C t^{-2\lambda} \lxr^{2\lambda +1} \|
v_\nl(\cdot)  \ldr^{2\lambda+1}\|_{L^1}   ,
\end{equation*}
where $D =  \int_\mbR \hphi_\nlo(x') v_\nl(x')dx'$.
It follows from Lemmas \ref{lemsxlxp}, \ref{lemllen},   and Proposition \ref{propcl} that
\begin{equation*}
 |S_2+S_3+S_4|\leq Ct^{-2\lambda }\lxr^{-3\lambda-1} \| \ldr^{3\lambda+1} v_\nl(\cdot)\|_{L^1}.
\end{equation*}
Combining the estimates for $I_i$ ($i=1, \cdots, 5$), we have
\begin{equation}\label{sinest}
\begin{aligned}
&\lxr^{-3\lambda-1} \left|- \frac{2}{\pi} \int_0^\infty \int_\mbR  \sin(t \s)\Im[G_\nl(x, x', \s) ]v_\nl(x') dx' d\s  -D  \hphi_\nlo(x) \right|\\
\leq  &C t^{-2\lambda} \| \ldr^{3\lambda+1} v_\nl(\cdot)
\|_{L^1}.
\end{aligned}
\end{equation}
The estimates \eqref{cosest} and \eqref{sinest} prove estimate \eqref{estsol1}.

This finishes the proof of Theorem \ref{thmZm}.
 \qed

%\newpage
\appendix
\section{Volterra Integral Equations}\label{secvol}
In this appendix, we state three useful lemmas on the existence and estimates for solutions of Volterra integral equations. This first one concerns existence and $L^\infty$ estimate. Its  proof can be found in \cite{Deift}.
\begin{lem}\label{lemvol}
Let  $g\in L^{\infty}(a,b)$. Suppose the integral kernel $K$ satisfies
\begin{equation*}
\mu=\int_a^{b}\sup_{x\in (a,y)}|K(x,y)|dy<\infty.
\end{equation*}
Then the Volterra integral equation
\begin{equation}
f(x)=g(x)+\int_x^{b}K(x,y)f(y)dy
\end{equation}\label{eqvolA}
has a unique solution satisfying
\begin{equation*}
\|f\|_{L^{\infty}(a,b)}\leq e^{\mu}\|g\|_{L^{\infty}(a,b)}.
\end{equation*}
Similarly, suppose that
\begin{equation*}
\nu=\int_a^{b}\sup_{x\in (y, b)}|K(x,y)|dy<\infty.
\end{equation*}
Then the Volterra equation
\begin{equation*}
f(x)=g(x)+\int_a^{x}K(x,y)f(y)dy
\end{equation*}
has  a unique solution satisfying
\begin{equation*}
\|f\|_{L^{\infty}(a, b)}\leq e^{\nu}\|g\|_{L^{\infty}(a, b)}.
\end{equation*}
\end{lem}

Finally, we have the following two lemmas on the derivatives of solutions for Volterra integral equations. The proofs can be found in \cite{SofferPrice}.
\begin{lem}\label{lemderivative}
If, in addition to the assumptions of  Lemma \ref{lemvol}, $g\in C^{\infty}(a, b)$ and the kernel $K$ is smooth in both variables on $(a, b)$ and satisfies
\begin{equation*}
\int_a^{b}|\partial_x^kK(x,y)|dy<\infty
\end{equation*}
for any $x\geq a $ and all $k\in\mbN$ then the solution $f$ is smooth on $(a, b)$. Furthermore, the derivatives of $f$ in \eqref{eqvolA} can be calculated by formal differentiation, i.e.,
\begin{equation*}
f^{(k)}(x)=g^{(k)}(x)-\sum_{j=0}^{k-1}(\kappa_jf)^{(k-1-j)}(x)+\int_x^{b}\partial_x^k K(x,y)f(y)dy
\end{equation*}
where $\kappa_j(x)=\partial_x^jK(x,y)|_{y=x}$.
\end{lem}

\begin{lem}\label{lemparameter}
Let $I\subset \mbR$ be open and suppose
\begin{equation*}
\int_a^{b}\sup_{x\in (a,y)}|\partial_{\s}^m K(x,y, \s)|dy<\infty
\end{equation*}
as well as $\partial_{\s}^m g(\cdot,\s)\in L^{\infty}(a, b)$ for all $m\in \mbN_0$ and $\s \in I$. Then the Volterra equation
\begin{equation*}
f(x,\s)=g(x,\s)+\int_x^{b}K(x,y,\s)f(y,\s) dy
\end{equation*}
has a unique solution $f(x,\s)$ for all $x\geq a$ and $\s \in I$ which is smooth in $\s$. Furthermore, we have $\partial_{\s}^m f(\cdot,\s)\in L^{\infty}(a, b)$ for all $m\in \mbN_0$ and the derivatives are given by
\begin{equation*}
\partial_{\s}^mf(x,\s)=\partial_{\s}^m g(x,\s)+\sum_{j=0}^m \left(\begin{array}{ll} m \\ j\end{array}\right)\int_x^{b} \partial_{\s}^j K(x,y,\s)\partial_{\s}^{m-j}f(y,\s)dy.
\end{equation*}
\end{lem}

\bigskip

 {\bf Acknowledgments:} Part of the work was done when Xie was visiting
The Institute of Mathematical Sciences, The Chinese University of
Hong Kong. He thanks  The Institute for its support and hospitality.

%\newpage

\end{document}